\documentclass[aps,prb,preprint,superscriptaddress,citeautoscript,nopacs]{revtex4-1}

\usepackage{amsmath}
\usepackage{amssymb}
\usepackage{bm}
\usepackage{epsfig}
\usepackage{comment}
\usepackage{hyperref}
\usepackage[usenames,dvipsnames]{xcolor}
\usepackage[normalem]{ulem}

\renewcommand{\v}[1]{{\bf #1}}
\newcommand{\bpm}{\begin{pmatrix}}
\newcommand{\epm}{\end{pmatrix}}

\newcommand{\nsection}[1]{\noindent{\textbf{#1}}\\ \noindent}

\begin{document}

             %
             %

%**********************************************************************************************************************

\title{Two-dimensional Peierls instability via zone boundary Dirac line nodes in layered perovskite oxides}

\author{Jin-Hong Park}
\thanks{equal contribution}
\affiliation{Center for Correlated Electron Systems, Institute for
  Basic Science (IBS), Seoul 08826, Korea}

\author{Seung Hun Lee}
\thanks{equal contribution}
\affiliation{Center for Correlated Electron Systems, Institute for
  Basic Science (IBS), Seoul 08826, Korea}
\affiliation{Department of Physics and Astronomy, Seoul National
  University, Seoul 08826, Korea}

\author{Choong H. Kim}
\affiliation{Center for Correlated Electron Systems, Institute for
  Basic Science (IBS), Seoul 08826, Korea}
\affiliation{Department of Physics and Astronomy, Seoul National
  University, Seoul 08826, Korea}

\author{Hosub Jin}
\email{[Corresponding author:]hsjin@unist.ac.kr}
\affiliation{Department of Physics, Ulsan National Institute of
  Science and Technology (UNIST), 50 UNIST, Ulsan 44919, South Korea}

\author{Bohm-Jung Yang}
\email{[Corresponding author:]bjyang@snu.ac.kr}
\affiliation{Center for Correlated Electron Systems, Institute for
  Basic Science (IBS), Seoul 08826, Korea}
\affiliation{Department of Physics and Astronomy, Seoul National
  University, Seoul 08826, Korea}
\affiliation{Center for Theoretical Physics (CTP), Seoul National
  University, Seoul 08826, Korea}

\date{\today}

\begin{abstract}
Interplay of Fermi surface topology and electron correlation is the quintessential ingredient underlying spontaneous symmetry breaking in itinerant electronic systems. In one-dimensional (1D) systems at half-filling, the inherent Fermi surface nesting makes the translationally invariant metallic state unstable, which is known as Peierls instability. Extending the scope of Peierls instability to two (2D) or three dimensions (3D), however, is not straightforward, since the Fermi surface in higher dimensions is generally not nested. In this work, we show that a perfectly nested Fermi surface can be realized in a class of 2D perovskite oxides, giving rise to 2D Peierls instability. Here the central role is played by the zone boundary Dirac line node (DLN) protected by two orthogonal glide mirrors induced by the rotation of oxygen octahedra. Especially, at a critical angle of the octahedron rotation, the zone-boundary DLN flattens, leading to logarithmically diverging susceptibility. We propose the 2D Peierls instability driven by dispersionless DLN as a principle mechanism for spontaneous symmetry breaking in various layered perovskite oxides including the antiferromagnetism of Sr$_2$IrO$_4$. As a clear signature of the 2D Peierls instability, we predict that the magnetic domain wall in Sr$_2$IrO$_4$ hosts localized soliton modes.
\end{abstract}

\maketitle

\begin{figure*}[t]
\centering
\includegraphics[width=16 cm]{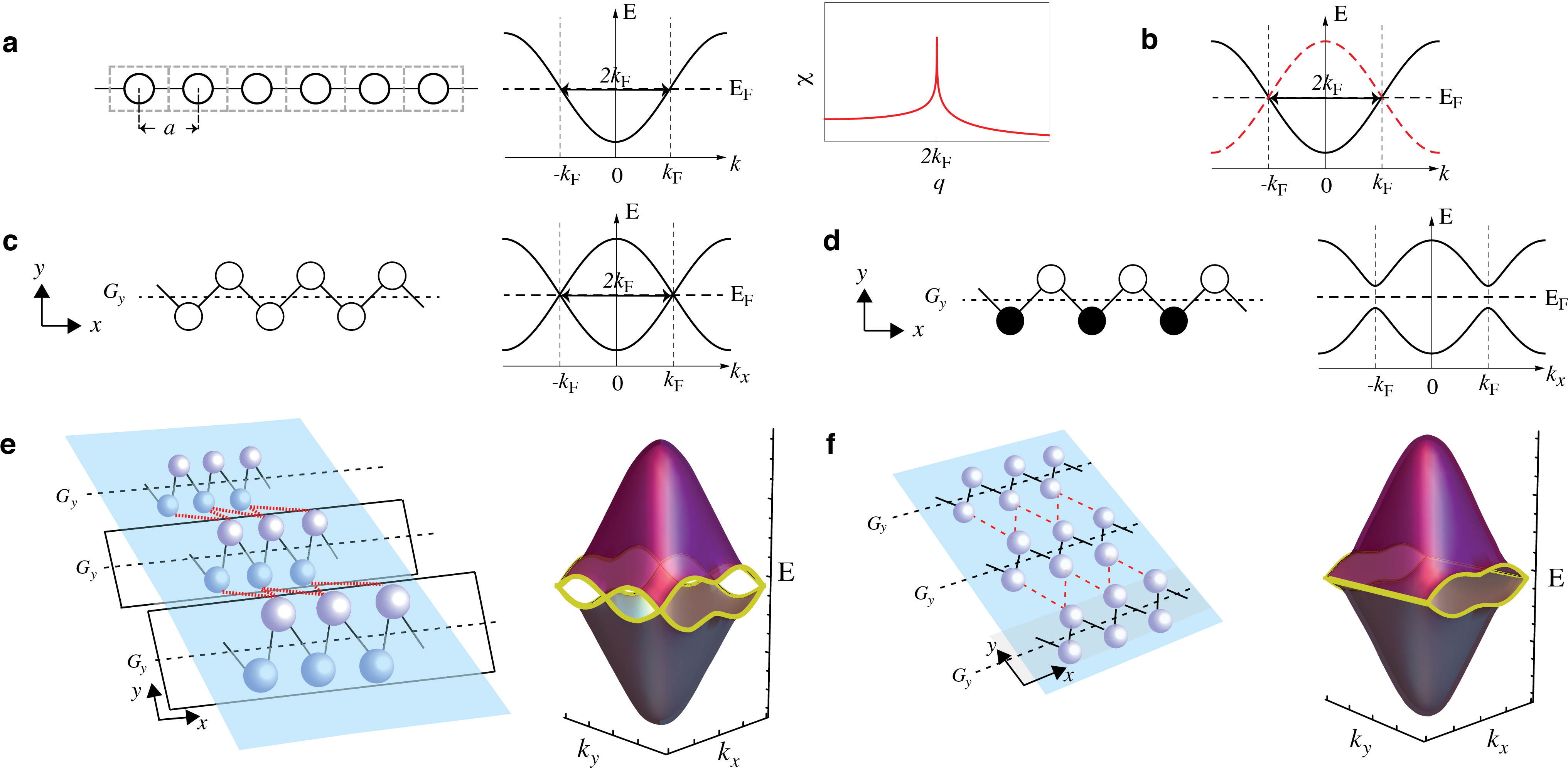}
\caption{
{\bf Peierls instability and glide mirror in one dimension (1D) and two dimensions (2D).}
({\bf a}) A monatomic chain having a single electron per site
has a nested half-filled band structure with the Fermi momentum $k_{F}$.
The relevant static susceptibility $\chi(q)$, which diverges logarithmically
at the wave vector $q=2k_{F}$.
({\bf b}) Band structure at the critical point after zone-folding.
({\bf c})
A zigzag shaped 1D chain having induced glide mirror symmetry and the resulting band structure.
Here the band degeneracy at the BZ boundary at $k=\pm k_{F}$ is protected by the
glide mirror induced by the lattice deformation.
({\bf d}) Consequence of glide symmetry breaking. Here the white and black dots indicate the two sites with different on-site potentials resulting from electron correlation.
({\bf e}) A 2D system composed of coupled 1D chains. A generic structure with a two-fold screw rotation can protect only a few Dirac points at the Brillouin zone (BZ) boundary.
({\bf f}) When the coupled 1D chains have an additional mirror symmetry about the 2D plane together with the original glide mirror, a Dirac line node appears along a BZ boundary
.
}\label{fig0}
\end{figure*}

Peierls instability is a ubiquitous mechanism originally suggested in a one-dimensional (1D) lattice at half-filling that leads to the spontaneous dimerization~\cite{peierls_quantum_1955}. Due to the inherent nesting of the 1D Fermi surface topology, the translationally invariant metallic state becomes unstable even in the presence of an infinitesimally weak interaction, manifested by the logarithmic divergence in its static susceptibility at the momentum $q=2k_{F}$ where $k_{F}$ indicates the Fermi momentum (Fig.~\ref{fig0}{\bf a}). In two (2D) or three dimensions (3D), however, the Fermi surface nesting is less likely, and the metallic state is stable as long as there is no effective attraction between electrons on the Fermi surface and the repulsive interaction between them is smaller than a certain threshold value~\cite{Shankar_1994}. Thus the interplay of Fermi surface topology and electron correlation lies at the heart of the weak coupling instability of the metallic state with translational invariance.

As an attempt to realize a 2D Peierls system, the laterally stacked 1D Peierls system can be constructed. For a 1D Peierls building block, let us consider a well-known polyacetylene chain at its critical point preserving the translation symmetry. In this system, Fermi surface nesting occurs in the form of a 1D Dirac point at the Brillouin zone (BZ) boundary~(Fig.~\ref{fig0}{\bf b}). By considering 1D polyacetylene chains at the critical point as being embedded in 2D, the 1D Dirac point can be extended to a flat Dirac line node (DLN) spanning the 2D BZ boundary. Therefore, the 2D extension of the Peierls instability is clued by the presence of dispersionless zone-boundary DLN at the critical point. In general, however, it is not easy to protect a line degeneracy in 2D systems, especially when both time-reversal $T$ and inversion $P$ symmetries exist together with spin-orbit coupling. In fact, even the zone boundary 1D Dirac point in a polyacetylene chain at its critical point is not a symmetry protected degeneracy but merely resulting from the unit cell doubling.

All those difficulties are remediable in the presence of nonsymmorphic crystalline symmetries such as glide mirrors or screw rotations, which is known to protect band degeneracies at the BZ boundary in general~\cite{book}. For instance, let us deform a straight 1D chain at its critical point to a zigzag form as shown in Fig.~\ref{fig0}{\bf c}. Due to the unit cell doubling, the deformed chain has a 1D Dirac point at the BZ boundary. Moreover, the deformation makes the zigzag chain invariant under a mirror or a two-fold rotation symmetry combined with a half-translation along the chain direction, that is, a glide mirror or a two-fold screw rotation symmetry is induced by the lattice deformation. Such an induced nonsymmorphic symmetry renders the zone boundary Dirac point symmetry-protected, thus it remains gapless as long as the corresponding nonsymmorphic symmetry is preserved (Fig.~\ref{fig0}{\bf d})~\cite{wieder_2016}. Arbitrary stacking of the zigzag-shaped chain does not guarantee a line degeneracy along the BZ boundary, since the combination of the two-fold screw rotation and inversion can at most protect the four-fold degeneracy only at a point (Fig.~\ref{fig0}{\bf e})~\cite{young_dirac_2015}. However, when the stacked chain system preserves the glide mirror of a 1D chain and has an additional in-plane mirror symmetry embracing the 2D plane, the four-fold degeneracy of the zone boundary DLN can remain intact even in the presence of spin-orbit coupling (Fig.~\ref{fig0}{\bf f}). Moreover, if the bandwidth of the symmetry-protected zone-boundary DLN can be controlled to become completely dispersionless, 2D Peierls instability can occur, leading to various symmetry breaking phenomena.

Here we show that such an intriguing idea can be realized in a wide class of layered perovskite oxides. The central role is played by the in-plane rotation of oxygen octahedra, which is a common lattice distortion among layered 2D perovskite oxides. It doubles the size of the unit cell and, at the same time, generates two orthogonal glide mirrors, leading to the DLN at the BZ boundary. Interestingly, the bandwidth of the nodal line dispersion can be controlled by changing the in-plane rotation angle $\theta$ of oxygen octahedra. When $\theta$  reaches a certain critical value $\theta_{c}$, the DLN on the BZ boundary becomes completely dispersionless, manifesting 2D Peierls instability with the logarithmically diverging susceptibility. We propose that the instability induced by the dispersionless zone boundary DLN is the principle mechanism for the canted antiferromagnetic ground state of Sr$_2$IrO$_4$. Given the magnetic ground state as a consequence of 2D Peierls instability, a magnetic domain wall (DW) of Sr$_2$IrO$_4$ is shown to host 1D localized soliton modes along the DW boundary. Since the origin of such a flat DLN is solely coming from the crystalline symmetry, we believe that the 2D Peierls instability can occur ubiquitously in various layered perovskite oxides sharing the same crystalline symmetry.

\subsection*{Results}

{\bf Lattice distortion induced nonsymmorphic symmetry.}

\begin{figure*}[t]
\centering
\includegraphics[width=16 cm]{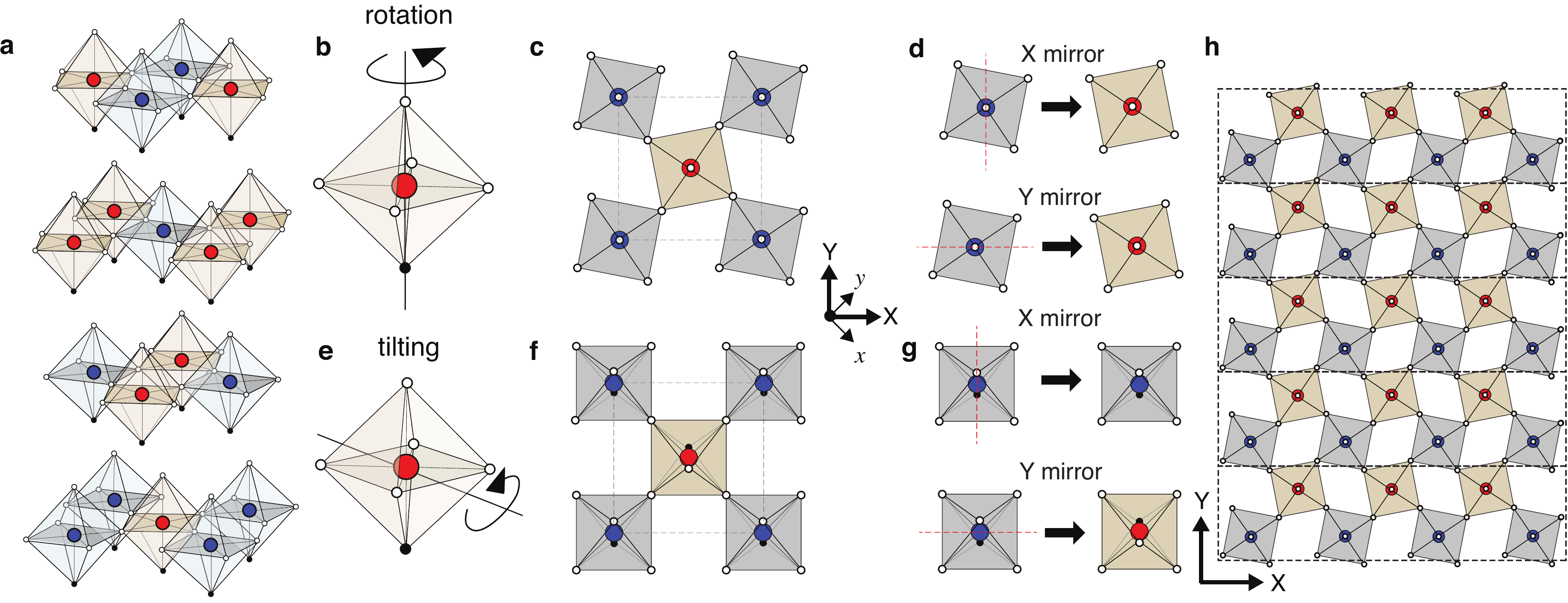}
\caption{
{\bf Rotation of oxygen octahedra and induced glide mirror in layered perovskite oxides.}
({\bf a}) Three-dimensional (3D) structure of a layered perovskite oxide.
({\bf b}) Rotation distortion of an octahedron due to the rotation about the $Z$-axis.
({\bf c}) Structure of a layer with the rotation distortion.
({\bf d}) Operation of the mirror symmetry about $YZ$ plane ($X$ mirror)and $XZ$ plane ($Y$ mirror) on a layer with rotation distortion.
({\bf e}) Tilting distortion of an octahedron due to the rotation about an in-plane axis.
({\bf f}) Structure of a layer with the tilting distortion.
({\bf g}) Operation of the mirror symmetry about $YZ$ and $XZ$ planes on a layer with the tilting distortion
({\bf h}) A two dimensional (2D) layer with rotation distortion can be considered as a coupled one dimensional (1D) zigzag chains with $G_{Y}$ symmetry stacked along the $Y$ direction.
}\label{fig:distortion}
\end{figure*}

Layered perovskite oxides with the chemical formula A$_{2}$BO$_{4}$, as shown in Fig.~\ref{fig:distortion}{\bf a}, normally undergo several kinds of structural distortions~\cite{braden_distortion_1998}. The most widely occurring distortions are the in-plane rotation of oxygen octahedra about the $z$-axis (rotation distortion, see Fig.~\ref{fig:distortion}{\bf b}) and another rotation of oxygen octahedra about an axis lying in the 2D plane (tilting distortion, see Fig.~\ref{fig:distortion}{\bf e}). Both rotation and tilting distortions double the size of the in-plane unit cell as shown in Fig.~\ref{fig:distortion}{\bf c,f}, and the relative orientation of the distorted octahedra between layers determines the overall space group symmetry of the 3D structure. In many cases, the bulk properties are mainly determined by the property of a monolayer due to the weak interlayer coupling.

There are several materials exhibiting rotation distortion \cite{carter_theory_2013, subramanian_sr$_2$rho$_4$_1994, ye_structure_2015, crawford_structural_1994, Yuan_From_2015}. For instance, Sr$_2$IrO$_4$ undergoes a rotation distortion of oxygen octahedra with the angle $\theta\sim 11^\circ$ in a staggered manner leading to the $\sqrt{2}\times\sqrt{2}$-type doubled unit cell before the antiferromangetic (AFM) ordering is developed~\cite{carter_theory_2013}. (See Fig.~\ref{fig:distortion}{\bf c}.) A similar distortion is observed in Ref. \onlinecite{subramanian_sr$_2$rho$_4$_1994, ye_structure_2015}  with $\theta\sim 9^\circ$. Such an in-plane rotation distortion changes the space group symmetry of the lattice from the symmorphic group $I4/mmm$ (no. 139) to the nonsymmorphic group $I4_1/acd$ (no. 142)~\cite{crawford_structural_1994} exhibiting two orthogonal glide mirrors (Fig.~\ref{fig:distortion}{\bf d}). Below we show that the nonsymmorphic symmetry induced by the rotation brings about remarkable physical consequences.

Explicitly, the two glide mirrors $G_{X,Y}\equiv\{M_{X,Y}|\frac{1}{2}\frac{1}{2}\}$ are the combination of an ordinary mirror $M_{X, Y}$ which inverts the sign of the $X$-or $Y$-coordinate and a partial translation $(\frac{1}{2}, \frac{1}{2})$ along the diagonal direction (See Fig.~\ref{fig:distortion}{\bf d}). Here we choose the $\sqrt{2}\times\sqrt{2}$-type doubled cell as a unit cell, and then the translations of the unit cell along the $X$ and $Y$ directions span the whole 2D lattice as shown in Fig.~\ref{fig:distortion}{\bf h}. The whole lattice can be viewed as a vertical stacking of horizontal zigzag chains analogous to Fig.~\ref{fig0}{\bf f}. The presence of these two orthogonal glide mirrors together with time-reversal $T$ and inversion $P$ guarantees the presence of a Dirac line node with four-fold degeneracy along the BZ boundary as explained in detail below.

Let us note that, in the case of the tilting distortion~\cite{rondinelli_structure_2011}, which exists in various materials including La$_2$CuO$_4$ and T-phase cuprates~\cite{axe_structural_1989,bozin_reconciliation_2015}, the distorted lattice hosts only one glide mirror as shown in Fig.~\ref{fig:distortion}{\bf g}, which can protect at most Dirac point nodes on the BZ boundary as shown in Ref.~\onlinecite{DiracSM_bjyang}. In this case, one cannot expect a significant enhancement of the susceptibility, thus we neglect the tilting distortion and focus on the rotation distortion in the forthcoming discussion.

{\bf Dirac line nodes (DLN) on the Brillouin zone boundary}

The two glide mirrors induced by the rotation distortion of oxygen octahedra can generate the four-fold degenerate DLN on the full BZ boundary due to the following reason. The point group symmetry of the system is generated by inversion $P$, and two glide mirrors $G_{X}$ and $G_{Y}$, which transform the spatial coordinate as
\begin{align}
P&:(X,Y)\rightarrow (-X,-Y),
\nonumber\\
G_{X}&:(X,Y)\rightarrow (-X+\frac{1}{2},Y+\frac{1}{2})\times i\sigma_{X},
\nonumber\\
G_{Y}&:(X,Y)\rightarrow (X+\frac{1}{2},-Y+\frac{1}{2})\times i\sigma_{Y},
\end{align}
where $\sigma_{X,Y,Z}$ indicate the spin Pauli matrices. By combining $P$ and $G_{X,Y}$, one can also define two two-fold screw rotations $S_{X}\equiv G_{X}P$ and $S_{Y}\equiv G_{Y}P$ and an in-plane mirror $M_{Z}\equiv G_{X}G_{Y}P$. In general, when $P$ and $T$ exist simultaneously, every band is doubly degenerate at each momentum. Due to the strong level repulsion between degenerate bands, it is not easy to achieve band crossing without proper additional symmetries~\cite{DiracSM_bjyang}, which in the present case are $G_{X}$ and $G_{Y}$.

Explicitly, let us first explain the role of $G_{Y}$ in protecting the band degeneracy along the BZ boundary, $k_{X}=\pm\pi$. As shown in Fig.~\ref{fig:distortion}{\bf h}, the distorted 2D lattice with rotation distortion can be considered as coupled 1D chains having $G_{Y}$. Since each chain hosts Dirac points at the BZ boundary with $k_{X}=\pm\pi$, the distorted 2D lattice can have a DLN along the BZ boundary with $\bm{k}=(\pm\pi,k_{Y})$ ($k_{Y}\in (-\pi,\pi)$). On the BZ boundary, the system is invariant under $PT$, $M_{Z}=G_{X}G_{Y}P$, and $S_{Y}=G_{Y}P=\{C_{2Y}|(\frac{1}{2},\frac{1}{2})\}$ where $C_{2Y}\equiv M_{Y}P$ is an ordinary two-fold rotation about the $Y$-axis. Let us note that $S_{Y}$ contains a half-translation along the $X$-direction perpendicular to its rotation axis. This indicates that the rotation axis of $S_{Y}$ is not located at the inversion center, that is, $S_{Y}$ is an off-centered two-fold rotation symmetry~\cite{offcentered_bjyang}. Because of such off-centered nature of $S_{Y}$, it anti-commutes with $PT$ on the BZ boundary
\begin{align}\label{eqn:anticommutation}
PTS_{Y}&=-e^{-ik_{Y}}S_{Y}PT,
\end{align}
which forces each Kramers pair on the BZ boundary to carry the same $S_{Y}$ eigenvalues, i.e., either $+ie^{ik_{Y}/2}$ or  $-ie^{ik_{Y}/2}$ (See Supplementary Information and Ref.~\onlinecite{offcentered_bjyang}). Then a DLN with four-fold degeneracy can occur, if two different Kramers pairs having distinct $S_{Y}$ eigenvalues are degenerate due to the presence of an additional symmetry. In fact, this is exactly the role played by $M_{Z}$ symmetry. Let us note that the spin orientation of $S_{Y}$ ($M_{Z}$) eigenstates is parallel to the $Y$($Z$)-axis since  $S_{Y}\propto i\sigma_{Y}$ ($M_{Z}\propto i\sigma_{Z}$) due to spin-orbit coupling. The orthogonal spin orientation between $S_{Y}$ and $M_{Z}$ eigenstates indicates the following anti-commutation relation
\begin{align}\label{eqn:commutation}
M_{Z}S_{Y}&=-S_{Y}M_{Z},
\end{align}
which, combined with Eq.~(\ref{eqn:anticommutation}), guarantees the four-fold degeneracy of the relevant DLN. The DLN on the BZ boundary $k_{Y}=\pm \pi$ can also be understood in a similar way. Therefore, the DLN spanning the full BZ boundary arises from the presence of two orthogonal glide mirrors in systems with $P$ and $T$ symmetries.

{\bf Tuning the bandwidth of the DLN via rotation distortion }

To demonstrate the presence of the DLN spanning the BZ boundary and how to control its bandwidth, we study a tight-binding Hamiltonian relevant to Sr$_2$IrO$_4$. Sr$_2$IrO$_4$ is a representative system in which the interplay of strong spin-orbit coupling and electron correlation can give rise to novel spin-orbit entangled ground states~\cite{kim_phase-sensitive_2009, moon_dimensionality-controlled_2008, kim_novel_2008,jin_anisotropic_2009, jackeli_mott_2009, wang_twisted_2011}. Since strong spin-orbit coupling splits 5$d$ $t_{2g}$  orbitals into a lower energy quartet and a higher energy doublet with the effective angular momentum $J_{\text{eff}}=3/2$ and $J_{\text{eff}}=1/2$, respectively, an Ir$^{4+}$ ion has a half-filled $J_{\text{eff}}=1/2$ state and fully-occupied  $J_{\text{eff}}=3/2$ states. Thus, the low energy band structure near the Fermi energy is dominated by the Ir $J_{\text{eff}}=1/2$ states, from which a lattice model Hamiltonian can be constructed.

\begin{figure*}[t]
\centering
\includegraphics[width=160mm]{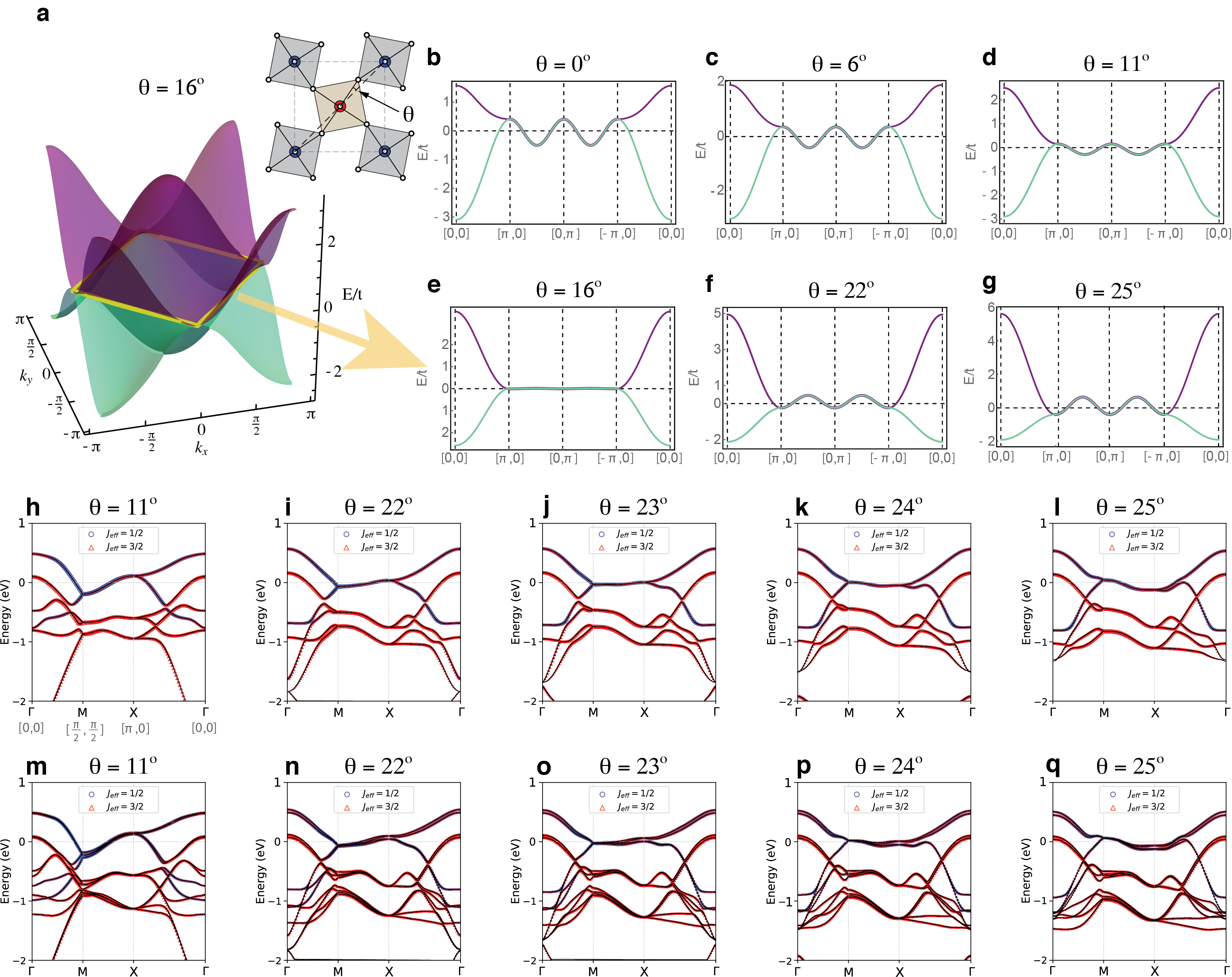}
\caption{
{\bf Dispersion of the Dirac line node (DLN) as a function of in-plane rotation angle $\theta$ in Sr$_2$IrO$_4$.}
The band structures in ({\bf a}-{\bf g}) are from tight-binding calculations while those in ({\bf h}-{\bf q}) are from first-principles calculations. 
({\bf a}) The paramagnetic band structure of a single Sr$_2$IrO$_4$ layer when $\theta = 16^\circ$. Here the DLN is dispersionless along the full Brillouin zone boundary.
The definition of the rotation angle $\theta$ relative to the undistorted lattice structure is also described.
({\bf b-g}) Dispersion of the DLN as $\theta$ varies. Here the purple and green lines are doubly degenerate. The DLN becomes completely flat at the critical angle $\theta\approx 16^\circ$.
({\bf h-l}) DFT band structures of a single Sr$_2$IrO$_4$ layer as $\theta$ varies while the Ir-O bond length is fixed. $J_{\textrm{eff}} =1/2$ (blue circle) and $J_{\textrm{eff}} =3/2$ (red triangle) bands are displayed by using different colors. The DLNs along the BZ boundary (M-X line) around the Fermi level become dispersionless at the critical angle $\theta\approx 23^\circ$. ({\bf m-q}) DFT band structures of the bulk Sr$_2$IrO$_4$ as $\theta$ varies while the Ir-O bond length is fixed. All symbols and colors are the same as in the case of a single Sr$_2$IrO$_4$ layer of ({\bf h-l}). The DLNs along the BZ boundary (M-X line) around the Fermi level become almost dispersionless at the critical angle $\theta\approx 23^\circ$. 
}\label{fig:energy-band}
\end{figure*}

The unit cell of Sr$_2$IrO$_4$ is composed of four layers of iridium oxide planes. For convenience, however, we first focus on the property of a single iridium oxide layer, and then include the influence of weak inter-layer coupling. By introducing $\psi^{\dag}(\v k)=[c^{\dag}_{A,\uparrow}(\v k),c^{\dag}_{A,\downarrow}(\v k),c^{\dag}_{B,\uparrow}(\v k),c^{\dag}_{B,\downarrow}(\v k)]$ as a basis, the lattice Hamiltonian for a single layer with a rotation distortion of an oxygen octahedron by an angle $\theta$ (see Fig.~\ref{fig:energy-band}{\bf a}) can be written as $\hat{H}_{\theta}=\sum_{\v k}\psi^{\dag}(\v k)H(\v k,\theta)\psi(\v k)$ in which
\begin{equation}
H(\v k, \theta) = \varepsilon_{1} (\v k, \theta) \sigma_0 \tau_x +  \varepsilon_{1d} (\v k, \theta) \sigma_z \tau_y + [\varepsilon_{2} (\v k, \theta)+ \varepsilon_{3} (\v k,
\theta)] \sigma_0 \tau_0
\label{eq:latticeH}
\end{equation}
where  $\varepsilon_{1,1d} (\v k, \theta)=2t_{1,1d}(\theta)[\cos(k_{x})+\cos(k_{y})]$, $\varepsilon_{2} (\v k, \theta)=4t_{2}(\theta)\cos k_{x}\cos k_{y}$, $\varepsilon_{3} (\v k, \theta)=2t_{3}(\theta)[\cos(2k_{x})+\cos(2k_{y})]$. Here we choose the unit translation vectors $\hat{x}$ and $\hat{y}$ of the undistorted lattice as a unit of real space coordinates for convenience. The explicit form of the hopping integral $t_{1,1d,2,3}(\theta)$ is shown in Methods. The Pauli matrices $\tau_{0,x,y,z}$ ($\sigma_{0,x,y,z}$) denote the $A$ and $B$ sublattice (the $J_{\mathrm{eff}} = 1/2$  pseudo-spin) degrees of freedom. The diagonal term $\varepsilon_{2} (\v k, \theta)$ ($\varepsilon_{3} (\v k, \theta)$) indicates the second (third) nearest neighbor hopping processes between the same sublattices with the same effective angular momenta. The $\theta$ dependence of the hopping integrals is derived from the Slater-Koster approximation~\cite{slater_simplified_1954}.

From Eq.~(\ref{eq:latticeH}), we have obtained the evolution of the band structures as a function of the rotation angle $\theta$, which is shown in Fig.~\ref{fig:energy-band}{\bf b-g}. The presence of the DLN spanning the full BZ boundary is clearly observed. The band structure of Sr$_2$IrO$_4$ with its rotation angle $\theta \sim 11^\circ$ matches well with the previously reported results~\cite{carter_theory_2013}. It is worthwhile to note that the overall bandwidth of the DLN on the BZ boundary strongly depends on $\theta$. Especially when $\theta \sim 16^\circ$, the DLN becomes completely flat as depicted in Fig.~\ref{fig:energy-band}{\bf a} and {\bf e}. Then the resulting semimetal with zone boundary DLN should be unstable even in the presence of an infinitesimally small interaction, which indeed links to the 2D Peierls instability.

The emergence of the flat DLN under rotation distortion is further supported by $\textit{ab-initio}$ density functional theory (DFT) calculations including spin-orbit coupling as shown in Fig.~\ref{fig:energy-band}{\bf h-q}. To observe the $\theta$-dependence in DFT band structure, the in-plane lattice constant is varied while the Ir-O bond length is fixed. Figure~3{\bf h-l} shows the evolution of DFT band structure for a single Sr$_2$IrO$_4$ layer.
During the variation of the rotation angle $\theta$, the bandwidth of the zone boundary DLN (M-X line) also changes, consistent with the tight-binding calculations. The four-fold degenerate DLN eventually becomes almost flat around the critical angle $\theta \sim 23^\circ$ as shown in Fig.~\ref{fig:energy-band}{\bf j}.
For the bulk Sr$_2$IrO$_4$ where the unit cell is composed of four monolayers, four distinct DLNs derived from $J_{\text{eff}}=1/2$ states on the BZ boundary are displayed in Fig.~\ref{fig:energy-band}{\bf m-q}. Since $G_X$, $G_Y$, $P$, $T$ symmetries are all preserved in the 3D structure, the fourfold degeneracy of each DLN is maintained.
The nearly degenerate DLNs along the BZ boundary (M-X line) around the Fermi level become almost dispersionless at the critical angle $\theta \sim 23^\circ$ as shown in Fig~\ref{fig:energy-band}{\bf o}. Although the critical angle predicted by the DFT calculations is not the same as that from the tight-binding calculations, the overall $\theta$-dependece of the zone boundary DNL indicates the consistency between them. To provide additional evidence for the tunability of the DLN via rotation distortion, we also have examined another type of $\theta$-variation, which is obtained by changing the Ir-O bond length while the in-plane lattice constant is fixed. One can again observe the emergence of the flat bands at a certain critical angle $\theta$ in both a monolayer and the bulk system, which supports the robustness of our theory on the band-width-controllable DNL. (For details, see Supplementary Information.)

{\bf Dispersionless DLN and the localized line states}

\begin{figure*}[t]
\centering
\includegraphics[width=160mm]{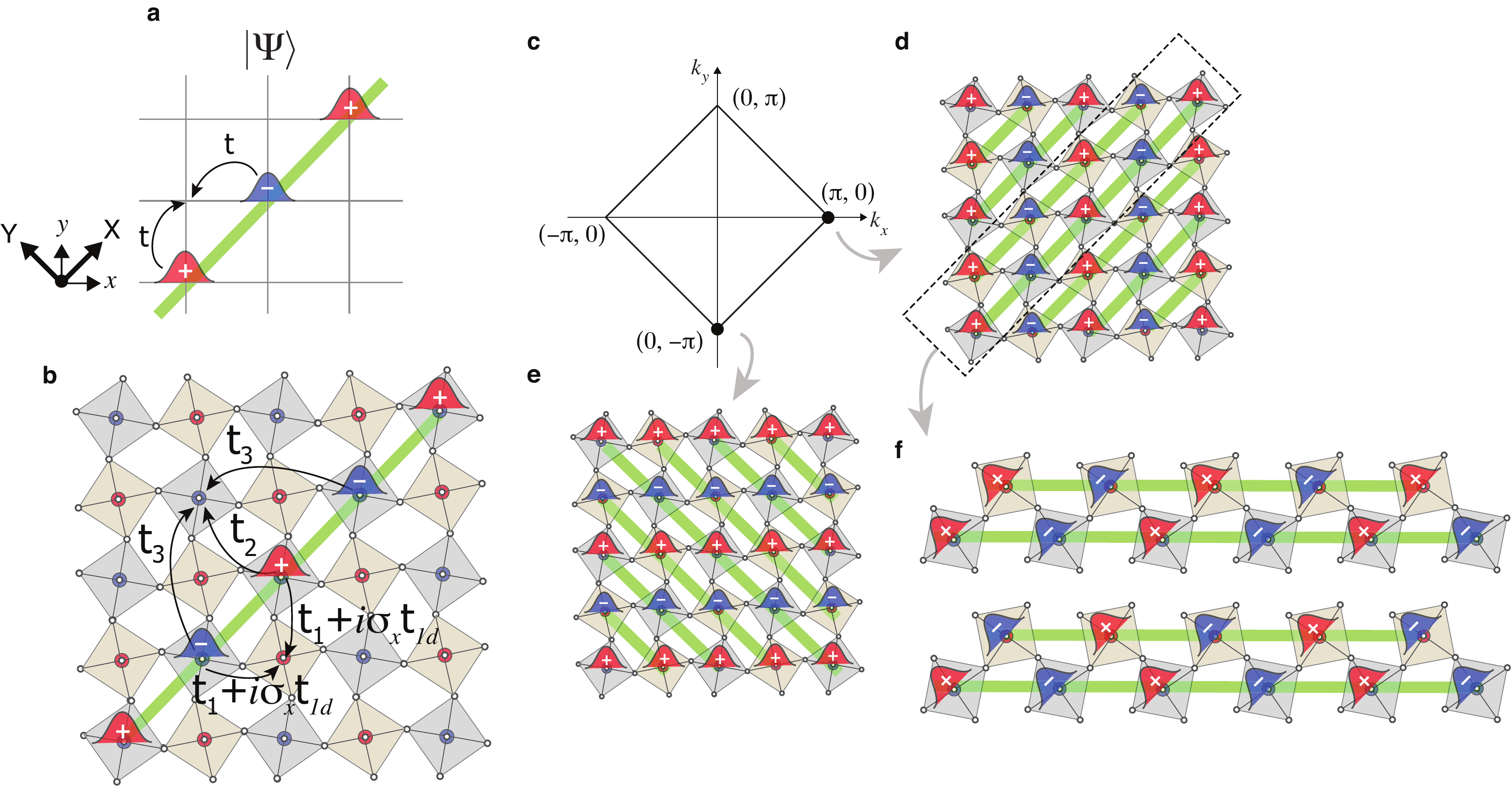}
\caption{
{\bf The origin of the flat Dirac line node (DLN) and the relevant localized line states.}
({\bf a}) Illustration of the origin of a localized line state. The hopping amplitudes  from two neighboring sites on a line to the common nearest neighbor site are canceled when the sign of the wave function alternates along the line.
({\bf b}) The hopping processes from a localized line state to its first, second, and third nearest neighbor sites.
({\bf c})The first Brillouin zone.
({\bf d}) The wave function of the flat DLN at the momentum $\bm{k}=(\pi,0)$.
({\bf e}) The wave function of the flat DLN at the momentum $\bm{k}=(0,-\pi)$.
({\bf f}) Two neighboring localized line states can form two degenerate eigenstates of a zigzag shaped chain with the momentum $\pi$. Including spin degrees of freedom, the localized line states can span four degenerate states with the momentum $\pi$.
 } \label{fig:chainstate}
\end{figure*}

When a band is completely dispersionless, it can be expressed as a linear combination of spatially localized eigenstates of the Hamiltonian. To fully account for the origin of the flat DLN on the BZ boundary, let us first consider a localized state shown in Fig.~\ref{fig:chainstate} defined along a diagonal line in the $2N\times2N$ square lattice as
\begin{equation}\label{eq:linestates}
\left|\Psi\right\rangle_{\ell_{\alpha},\sigma}^{\alpha}=\frac{1}{\sqrt{2N}}\sum_{\bm{r}\in \ell_{\alpha}}(-1)^{r_{x}}c_{\sigma}^{\dagger}(\bm{r})\left|0\right\rangle,
\end{equation}
where $\ell_{\alpha=p,n}=1, 2, ..., 2N$ is the labeling for a diagonal line with positive ($\alpha=p$) or negative ($\alpha=n$) slope, while a diagonal line with odd (even) $\ell_{\alpha}$ is composed of sites belonging to the A(B)-sublattice. $\bm{r}=(r_{x},r_{y})$ indicates the coordinate of a lattice site, and $\sigma=\pm$ denotes the effective angular momentum $J_{\text{eff},z}=\pm 1/2$. Basically, $\left|\Psi\right\rangle_{\ell_{\alpha},\sigma}^{\alpha}$ represents a line of states whose local wave function amplitude changes the sign alternatively along the line. Illustrations of such diagonal line states with positive and negative slopes are shown in Fig.~\ref{fig:chainstate}{\bf d,e}.

A strictly localized wave function can be an eigenstate of the Hamiltonian only when the sum of hopping amplitudes onto sites outside the support of the wave function vanishes~\cite{bergman_band_2008}. To examine the condition for $\left|\Psi\right\rangle_{\ell_{\alpha},\sigma}^{\alpha}$ to be an eigenstate of the Hamiltonian, we first consider the hopping processes between nearest-neighbor sites described by the the following Hamiltonian
\begin{equation}
\hat{H}_{1}=t_{1}(\theta)\sum_{\langle \bm{r},\bm{r'}\rangle,\sigma}\left[c_{\sigma}^{\dagger}(\bm{r})c_{\sigma}(\bm{r'})+h.c.\right]+t_{1d}(\theta)\sum_{\langle \bm{r},\bm{r'}\rangle,\sigma}\sigma \left[i c_{\sigma}^{\dagger}(\bm{r})c_{\sigma}(\bm{r'}) + h.c.\right],
\end{equation}
where $\langle \bm{r},\bm{r'}\rangle$ denotes a pair of nearest-neighbor sites belonging to different sublattices located at $\bm{r}$ and $\bm{r'}$, respectively. In momentum space, $\hat{H}_{1}$ gives rise to the terms $\varepsilon_{1/1d} (\v k, \theta)=2t_{1/1d}(\theta)[\cos(k_{x})+\cos(k_{y})]$ in Eq.~\eqref{eq:latticeH}. By applying $\hat{H}_{1}$ to $\left|\Psi\right\rangle_{\ell_{\alpha},\sigma}^{\alpha}$, one can easily find that $\hat{H}_{1}\left|\Psi\right\rangle_{\ell_{\alpha},\sigma}^{\alpha}=0$. Namely, due to the alternating sign of the wave function along the line, the hopping amplitudes to neighboring sites are canceled (see Fig.~\ref{fig:chainstate}{\bf a}), thus $\left|\Psi\right\rangle_{\ell_{\alpha},\sigma}^{\alpha}$ becomes a localized eigenstate with zero energy. Therefore the diagonal line states $\{\left|\Psi\right\rangle_{\ell_{\alpha},\sigma}^{\alpha}\}$ form a set of $8N$ independent and degenerate localized eigenstates.

Now we construct momentum eigenstates by taking a suitable linear combination of the localized diagonal line states as follows
\begin{align}
\left|\Psi\right\rangle_{A,\sigma}^{\alpha}(\phi)&=\frac{1}{\sqrt{N}}\sum_{m=1}^{N}e^{i 2m\phi}\left|\Psi\right\rangle_{\ell_{\alpha}=2m,\sigma}^{\alpha},
\nonumber\\
\left|\Psi\right\rangle_{B,\sigma}^{\alpha}(\phi)&=\frac{1}{\sqrt{N}}\sum_{m=1}^{N}e^{i (2m-1)\phi}\left|\Psi\right\rangle_{\ell_{\alpha}=(2m-1),\sigma}^{\alpha}.
\end{align}
As shown in Methods, it is straightforward that $\left|\Psi\right\rangle_{A,\sigma}^{p}(\phi)$ is a plane wave state with momentum $\bm{k}$ satisfying $k_{x}+k_{y}=\pi$. Likewise, one can check that $\left|\Psi\right\rangle_{B,\sigma}^{p}(\phi)$ is another plane wave state with the same momentum. Taking the pseudo-spin $\sigma$ into account, we have found four linearly independent degenerate eigenstates which are dispersionless along the BZ boundary satisfying $k_{x}+k_{y}=\pi$. By repeating similar procedures, one can also show that $\{\left|\Psi\right\rangle_{A/B,\sigma}^{n}(\phi)\}$ form four-fold degenerate eigenstates which are dispersionless along another BZ boundary satisfying $k_{x}-k_{y}=\pi$.

When the hopping processes between the second and third nearest-neighbor sites are included, the diagonal line states become dispersive. Thus dispersionless DLN can be spanned by the diagonal line states only under a certain limited condition, which, in the present problem, corresponds to the case when the rotation angle of an oxygen octahedron reaches the critical value $\theta_{c} \sim 16^\circ$. The Hamiltonian describing the hopping amplitudes between the second ($t_{2}$) and the third ($t_{3}$) nearest-neighbor sites is given by
\begin{equation}
\hat{H}_{23}=t_{2}\sum_{\langle\langle \bm{r},\bm{r'}\rangle\rangle,\sigma}\left[c_{\sigma}(\bm{r})^{\dagger}c_{\sigma}(\bm{r'})+h.c.\right]+t_{3}\sum_{\langle\langle\langle \bm{r},\bm{r'}\rangle\rangle\rangle,\sigma}\left[c_{\sigma}^{\dagger}(\bm{r})c_{\sigma}(\bm{r'})+h.c.\right],
\end{equation}
which, in momentum space, gives rise to $\varepsilon_{2}(\textbf{k},\theta)$ and $\varepsilon_{3}(\textbf{k},\theta)$ in Eq.~(\ref{eq:latticeH}).
By applying $\hat{H}_{23}$ to $\left|\Psi\right\rangle_{\ell_{\alpha},\sigma}^{\alpha}$, we obtain
\begin{align}
\hat{H}_{23}\left|\Psi\right\rangle_{\ell_{\alpha},\sigma}^{\alpha}&
=(2t_{3}-t_{2})[\left|\Psi\right\rangle_{\ell_{\alpha}+2,\sigma}^{\alpha}+\left|\Psi\right\rangle_{\ell_{\alpha}-2,\sigma}^{\alpha}]-2t_{2}\left|\Psi\right\rangle_{\ell_{\alpha},\sigma}^{\alpha}.
\end{align}
Thus, for $\left|\Psi\right\rangle_{\ell_{\alpha},\sigma}^{\alpha}$ to be the eigenstate of $\hat{H}_{23}$, the condition $t_{2}=2t_{3}$ should be satisfied. In fact, this is an identical condition to obtain $\theta_{c}$ at which the dispersionless DLN appears on the BZ boundary. When further neighbor hopping processes are included additionally, diagonal line states may not be localized eigenstates anymore, but the suitable linear combination of them can recover compact localized states spanning a flat zone boundary DLN. Therefore a zone boundary DLN is generally expected to exist as along as the symmetry of the system is maintained and there are enough number of control parameters such as the rotation angle of oxygen octahedra.

It is worthwhile to note that a pair of neighboring diagonal line states can construct the degenerate eigenstates of a zigzag-shaped chain shown in Fig.~\ref{fig:chainstate}{\bf f}, which consist of the aforementioned zone boundary Dirac point of the 1D Peierls system. This again supports the idea of viewing the distorted 2D lattice as stacking 1D zigzag-shaped chains with glide symmetry.

{\bf Magnetic instability driven by the dispersionless DLN}

Here we discuss the physical consequence induced by the dispersionless DLN on the BZ boundary. Let us note that in a polyacetylene chain with zigzag-type deformation at its critical point, the static susceptibility diverges logarithmically at the momentum $q=0$ (modulo a reciprocal lattice vector) signaling a sublattice symmetry breaking. Although the nature of the resulting ground state depends on the effective interaction, the sublattice symmetry breaking always accompanies the breaking of the glide mirror that otherwise protects the Dirac point, leading to a gapped phase with lower energy. A similar idea can be applied to a 2D Peierls system driven by a flat zone boundary DLN. Due to the perfect Fermi surface nesting from the dispersionless DLN, the uniform static susceptibility with the momentum $\bm{q}=0$ (modulo a reciprocal lattice vector) diverges logarithmically. An order parameter breaking the glide mirror symmetry can lift the degeneracy of DLN leading to a gapped insulator with lower energy. In the case of Sr$_2$IrO$_4$, its ground state is known to be a Neel-type AFM with in-plane spin canting (in-plane canted AFM). In the following, we examine the magnetic instability of this system focusing on the role of the zone boundary DLN whose bandwidth can be controlled by varying the rotation angle $\theta$ of oxygen octahedra.

Previous theoretical studies have shown that the lattice model for a monolayer composed of $J_{\textrm{eff}}=1/2$ states cannot capture the spin anisotropy of the system~\cite{carter_theory_2013,wang_twisted_2011}. Thus, to obtain the in-plane canted AFM ground state numerically, we consider the 3D structure with the unit cell comprised of four layers. Let us note that, as long as the glide symmetries are preserved, the almost flat DLN can still appear even in the presence of inter-layer coupling, which merely renormalizes the critical angle at which the zone boundary DLN becomes dispersionless.
We determine the magnetic ground state derived from the DLN and the relevant phase diagram by studying both the RPA-type spin susceptibility and the self-consistent mean field theory.

The general form of the spin susceptibility is given by
$\chi^{ij}_{\alpha \alpha',l l'} (\v q) =-\int_{0}^\beta d \tau \langle S^i_{\alpha l} (\v q, \tau)S^j_{\alpha' l'} (- \v q, 0) \rangle$ where the spin
operator is defined as $S^i_{\alpha l} (\v q, \tau) = \sum_{\v p}
c^{\dag}_{\v p, \alpha l}(\tau) [\sigma^i] c_{\v p + \v q, \alpha
  l}(\tau)$. Here $\alpha, \alpha'$ and $l,
l'$ indicate the sublattice and layer indices, respectively. To distinguish the two candidate ground states, the in-plane canted AFM and the $c$-axis collinear AFM, we have computed the spin susceptibility $\chi^{+-}_{\textrm{AFM}} (\v q)$ and
$\chi^{zz}_{\textrm{AFM}} (\v q)$ at the momentum $\v q$ considering the staggered spin operator $S'^i= S_{A}^i - S_{B}^i$ in the unit cell. As shown in Fig.~\ref{fig:phasediagram}{\bf b}, the spin susceptibility develops a peak at $\v q = (0,0)$. The magnitude of the spin susceptibility for in-plane AFM ordering is larger than that of $c$-axis AFM ordering as indicated in Fig.~\ref{fig:phasediagram}{\bf c}, which agrees with the experimental results~\cite{kim_phase-sensitive_2009}.
Upon varying the rotation angle of IrO$_6$ octahedron, the susceptibility at $\v q = 0$ rapidly grows and reaches its maximum at a critical angle where energy spectrum along the BZ boundary becomes almost flat (Fig.~\ref{fig:phasediagram}{\bf d}). Using the RPA-corrected spin susceptibility $\chi^{\textrm{RPA}} = \frac{\chi^0}{1- U \chi^0}$, we can determine the critical value of the Coulomb interaction $U_{c}$ from the condition that $\chi^{\textrm{RPA}}$ diverges at $U=U_{c}$, which is summarized in the phase diagram shown in Fig.~\ref{fig:phasediagram}{\bf e}.

Additionally, to confirm the magnetic ordering pattern suggested by the spin susceptibility, we have performed a self-consistent mean field calculation of a Hubbard-type model Hamiltonian with on-site repulsion : $H = H_t + H_U$ where $H_t$ is a $16 \times 16$ tight-binding Hamiltonian including the sublattice, $J_{\textrm{eff}}=1/2$ pseudo-spins, and the layer degrees of freedom.  The mean-field decoupling of the Hubbard interaction is implemented as $H_U = U \sum_i n_{i \uparrow} n_{i \downarrow} \rightarrow -U \sum_i (2 \langle \v S_i \rangle \cdot \v S_i - \langle \v S_i\rangle^2)$ with $\langle\v S_i \rangle= \langle\sum_{\sigma, \sigma'} c^{\dag}_{i\sigma}{{\bm \sigma}_{\sigma \sigma'} \over 2} c_{i \sigma'}\rangle\equiv \bm{m}_{i}$.  We determine the magnetic ordering pattern by computing the order parameter $\v m^A =(m^A_x, m^A_y, m^A_z)$ for sublattice A and $\v m^B = (m^B_x, m^B_y,m^B_z)$ for sublattice B in the bottom layer self-consistently.
Adopting the ``up-down-down-up" type interlayer spin ordering pattern confirmed in previous studies~\cite{carter_theory_2013,upupdndn_jkim}, the order parameters in the other three layers are chosen accordingly.  The resulting phase diagram is demonstrated in Fig.~\ref{fig:phasediagram}{\bf e}. The mean field theory shows that the critical interaction strength $U$ at which the phase transition occurs, becomes minimal when the rotation angle reaches the critical value $\theta = 16^\circ$. As shown in Fig.~\ref{fig:phasediagram}{\bf e}, $U_{c}$ determined from $\chi^{\textrm{RPA}}$ agrees well with the result from the self-consistent mean field theory, which confirms that the in-plane canted AFM ground state in Sr$_2$IrO$_4$ manifests itself as a consequence of 2D Peierls instability.

Let us note that in Ba$_2$IrO$_4$ where Sr is replaced by Ba, the ground state is an AFM insulator although there is no rotation distortion $(\theta = 0)$. In this case, since the space group of the system without rotational distortion is symmorphic, one may expect that our theory based on the zone boundary Dirac line node cannot be applied. However, let us stress that this is not the case. If we plot the band structure by using the same doubled unit cell, one can still observe the zone boundary Dirac line node, and the magnetic instability of the system can still be described by using the same theoretical framework. The physical property of the system is independent of the unit cell choice. The existence of the zone boundary DLN in Ba$_2$IrO$_4$ is confirmed by the tight-binding approach (see Fig. 3{\bf b}) as well as DFT calculations where $\sqrt{2}\times \sqrt{2}$ unit cell is used (see Supplementary Information). In particular, one can also observe the flattening of the DLN as $\theta$ is increased artificially, although the real system with $\theta=0$ can develop AFM state due to the relatively strong $U$. This clearly shows that the Dirac line node based AFM mechanism is still valid in Ba$_2$IrO$_4$ system as well.

{\bf Domain wall solitons.}

The emergence of zero dimensional (0D) soliton modes localized at a domain wall (DW) is a hallmark of 1D Peierls systems, which is normally described by the Su-Schrieffer-Heeger (SSH) model~\cite{ssh}. As a natural extension, in 2D Peierls systems, one can expect emerging 1D soliton modes localized along a DW, which can be considered as the coupled 0D soliton modes stacked along the DW direction. To demonstrate this idea, we have studied the energy spectrum of a configuration at the critical rotation angle with a magnetic DW between two canted AFM domains with the net ferromagnetic moments along the $+Y$ and $-Y$ directions, respectively, as shown in Fig.~\ref{fig:DW}{\bf a}. For simplicity, we first have considered a “smooth wall” in which the magnitudes of local magnetic moments are smoothly scaled down to zero as we approach the DW from the bulk region whereas the direction of spins in each domain is fixed. As shown in Fig.~\ref{fig:DW}{\bf b,c}, one can clearly observe two in-gap states as the 1D soliton modes localized along the DW. When the local magnetic moment at the DW ($m_{\text{DW}}$) is zero, the in-gap states appear exactly at the zero energy, which are dispersionless due to the same reason as the localized chain states appear. On the other hand, as $m_{\text{DW}}$ increases, the two in-gap states couple and develop dispersion with a small gap between them. However, even when $m_{\text{DW}}$ becomes as big as the local magnetic moment in the bulk, the gap between the soliton modes is ten times smaller than the bulk gap as shown in Fig.~\ref{fig:DW}{\bf d}. Both the lattice model study and the low energy effective Hamiltonian analysis consistently show that the in-gap states localized at the DW share the same origin as the DW soliton predicted in the original SSH model as discussed in detail in Supplementary Information.

To confirm the robustness of the in-gap states independent of the detailed structure around the DW, we have studied the energy spectra of various DW configurations considering different DW direction and changing the orientation of the net ferromagnetic moment as shown in Fig.~\ref{fig:DW}{\bf e-g}. For instance, allowing the rotation of spin directions around the DW, we have considered the Neel-type and Bloch-type DWs, both of which possess similar in-gap states. (See Supplementary Information.) When the DW is parallel to either the $x$ or $y$ direction, the in-gap states appear more dispersive as compared to the case of DWs parallel to the $X$ or $Y$ directions as shown in Fig.~\ref{fig:DW}{\bf f,g}. In all cases, it is found that the in-gap states localized at the DW are robust and well-separated from the bulk states, thus they are detectable through local conductivity measurements~\cite{metallic_DW_ma, topological_order_Cheong}.

\subsection*{Discussion}

We conclude with the discussion about experimental evidence supporting the presence of zone boundary DLN. According to the recent ARPES study of La doped iridates La$_x$Sr$_{2-x}$IrO$_4$, a collapse of the charge gap due to electron doping results in a paramagnetic metallic state with nodal fermionic excitations~\cite{Ladoped_iridate_exp}. Since electron doping shifts the position of the Fermi level, which weakens the instability associated with the DLN, it is natural to expect the recovery of the zone boundary DLN as long as two orthogonal glide symmetries remain intact in the doped paramagnetic state. If one of the two glide mirrors is broken, for instance due to the presence of another nonmagnetic order parameter, the zone boundary DLN can be deformed to Dirac points protected by the remaining glide mirror~\cite{hiddenorder_iridate_theory}. Overall, the relatively weak dispersion of the zone boundary DLN in Sr$_{2}$IrO$_4$ makes the critical interaction $U_{ c}$ small, thus the recovery of the paramagnetic semimetal with DLN requires a huge reduction of the effective Coulomb repulsion through carrier doping~\cite{Ladoped_iridate_theory}.

On the other hand, in Sr$_{2}$RhO$_4$ where Ir$^{4+}$ is replaced by Rh$^{4+}$ having five valence electrons in $4d$ orbitals, a paramagnetic metallic state is realized due to the weak electron correlation and large effective bandwidth. Previous ARPES study and first-principles calculation consistently show the presence of zone boundary DLN~\cite{Sr2RhO4_exp1,Sr2RhO4_exp2,Sr2RhO4_theory}. To induce an instability by controlling the rotation angle of RhO$_6$ octahedra, either by applying electric field or chemical doping would be an intriguing topic for future studies. By means of DFT calculations, it can be shown that the bandwidth of the zone boundary DLN in Sr$_2$RhO$_4$ also changes as a function of the rotational angle. It is minimized by a suitable choice of the rotational angle as shown in supplementary information. It is worthwhile to mention that the position of the DLN is deviated from the Fermi level due to the overlap with other dispersive bands resulting in a large effective bandwidth in total. As a consequence, the instability of the DLN is  compromised and the system remains the paramagnetic metallic state in Sr$_2$RhO$_4$. 

The recent second harmonic generation study as well as the neutron diffraction measurements~  \cite{Feng_Neutron_2013, Torchinsky_SHG_2015, Feng_Neutron_2015} indicate that the crystal structure of Sr$_2$IrO$_4$ is described by space group $I4_1/a$, which is different from the nonsymmorphic group $I4_1/acd$. The modification of the crystal structure is associated with the staggered tetragonal distortion of oxygen octahedron such that the ratio of the out of plane Ir-O bond length and the in-plane Ir-O bond length at the two Ir sublattice sites are different by 0.1 percent. Even though such a small tetragonal distortion is enough to generate superlattice peaks for structure analysis, it hardly affects the electronic structure and thus the instability of the DLN as well. The DFT band structure calculations shows that the energy splitting due to the staggered tetragonal distortion is indeed negligible. (For details, see Supplementary Information in the section of ``DFT band structure calculations including staggered tetragonal distortion".)

Finally, let us note that our theory can help resolve the controversy about the origin of the AFM in Sr$_2$IrO$_4$, which is typically ascribed either to the Slater mechanism or to the Mott mechanism~\cite{arita2014mott}. The in-plane AFM ordering doubles the unit cell and it is accompanied by the insulating behavior, supporting the Slater mechanism. On the other hand, the fact that the unit-cell doubling happens above the Neel temperature and the insulating behavior is accompanied by significant band renormalization supports the Mott mechanism. According to our theory, the correct way to describe the AFM is to take into account both viewpoints at the same time. Namely, the doubling of the unit cell due to the lattice distortion generates symmetry protected zone boundary DLNs which provide a platform for magnetic instability. Then subsequent flattening of the DLN enhances the effect of Mott correlation, which eventually drives the AFM ground state. We believe that our theory reveals a clear microscopic picture to understand the interplay between the symmetry protected band structure and the Mott correlation, leading to the AFM ground state in Sr$_2$IrO$_4$. The intricate balance among the spin-orbit coupled band structure, lattice symmetry, and electron correlation underlies the magnetic instability of Sr$_2$IrO$_4$, which would provide a new perspective to envision various the spin-orbit coupled complex correlated electron systems in general.

\noindent
\textbf{Methods}

\nsection
{Details of hopping integrals in tight-binding Hamiltonian}
The bandwidth control by varying the rotational angle of octahedron is important to examine the nesting induced instability in layered perovskite oxide systems. Here we explain how the hopping integrals in the tight-binding Hamiltonian in Eq.~\eqref{eq:latticeH} are obtained. We use the Slater-Koster methods to derive the $\theta$-dependent hopping integrals between spin-orbit coupled states of $ | J_{\mathrm{eff}} = \pm \frac{1}{2} \rangle = \frac{1}{\sqrt{3}} (|d_{yz} \mp s \rangle \pm i |d_{zx} \mp s \rangle \pm |d_{xy} \pm s \rangle)$, where $s$ refers to the spin. It is based on the idea that the hopping integrals can be decomposed into several hopping elements such as $V_{dd\pi}, V_{dd\delta},  V_{dd\sigma}$ in the $d$-orbitals basis and can be parameterized with respect to the relative displacement between two orbitals. The relative displacement is then adjusted by the amount of the angle for rotational distortion $\theta$. (See also Slater-Koster parameter method in Supplementary Information).
Accordingly, the explicit form of hopping integrals in \eqref{eq:latticeH} are
\begin{align}
  2t_{1} &= \frac{1}{12} [12V_{dd \pi} -V_{dd \delta} - 3V_{dd \sigma} ]  +\frac{2}{3}[V_{dd\pi} +V_{dd\delta}] \cos 2\theta - \frac{1}{12} [4V_{dd \pi} - V_{dd \delta} - 3 V_{dd \sigma}] \cos 4\theta,\nonumber\\
2t_{1d} &= \frac{2}{3} [V_{dd \pi} + V_{dd \delta}] \sin 2 \theta,\nonumber\\
  4t_{2} &= \frac{1}{2}[4V_{dd\pi} + 3V_{dd\delta}+V_{dd\sigma}]-\frac{1}{6} [4 V_{dd\pi} - V_{dd\delta} -3 V_{dd\sigma} ] \cos 4\theta, \nonumber\\
  2t_{3} &= \frac{1}{4}[4V_{dd\pi n} + 3V_{dd\delta n} +V_{dd\sigma n} ] +\frac{1}{12} [4 V_{dd \pi n} - V_{dd \delta n} -3 V_{dd \sigma n}] \cos 4\theta.
\end{align}
The rotational angle dependence $\sim\cos 4\theta$ results from the intra-orbital hybridization between $d_{xy}$-orbitals. The rotational angle dependence $\sim\cos 2\theta$ in $t_{1}$ describes the intra-orbital hybridization within $d_{yz}$-orbitals or $d_{zx}$-orbitals whereas the rotational angle dependence $\sim\sin 2\theta$ in $t_{1d}$ describes the $inter$-orbital hybridization between $d_{yz}$-orbitals and $d_{zx}$-orbitals. The hopping elements between nearest neighbor sites are chosen as $(V_{dd\pi}, V_{dd\delta} , V_{dd\sigma} ) = (1, -0.25, -1.5)$ and those for the next nearest neighbor sites are $(V_{dd\pi n}, V_{dd\delta n}, V_{dd\sigma n} ) = l \times (V_{dd\pi}, V_{dd\delta}, V_{dd\sigma})$ with $l = 0.07$. The factor $l$ reflects the reduction of hopping integral with respect to the distance. Naively, the reduction factor has to be chosen as $l = (1/2)^5 \approx 0.0312$, but considering the results from \textit{ab initio} calculations~\cite{carter_theory_2013}, we have used $l = 0.07$.

\nsection
{First-principles calculations}
Our electronic structure calculations were based on density-functional theory within the local density approximation (LDA) as implemented in Elk code~\cite{elk-website}.
For the exchange-correlation energy part of the LDA functional, we used the Perdew-Zunger parameterization of the Ceperly-Alder data~\cite{Accurate_Perdew}.
Spin-orbit coupling (SOC) was included in the second-variational scheme.
Brillouin zone integrations were performed using 6$\times$6$\times$3 grid sampling during the self-consistent calculations.

\begin{figure*}[t]
\centering
\includegraphics[width=160mm]{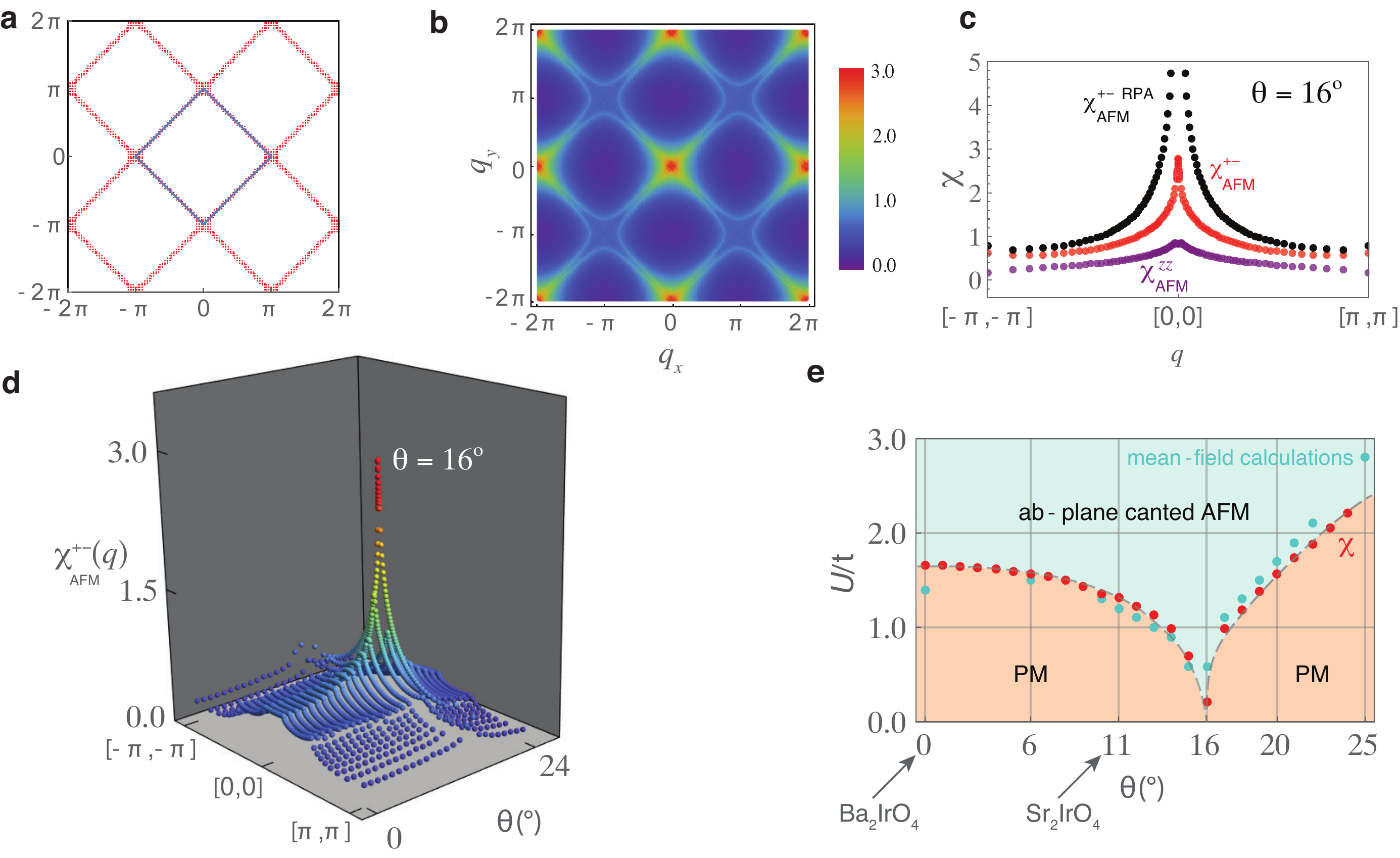}
\caption{
{\bf Static susceptibility at the critical angle and the generic phase diagram.}
({\bf a}) The Fermi surface when the dispersion of the Dirac line node (DLN) along the Brillouin zone (BZ) boundary becomes flat.
({\bf b}) Susceptibility $\chi^{+-}_{\mathrm{AFM}}(\bm{q})$ has a sharp peak at $\v q = (0,0)$ modulo a reciprocal lattice vector.
({\bf c}) The comparison of $\chi^{+-}_{\mathrm{AFM}}$ and $\chi^{zz}_{\mathrm{AFM}}$. The susceptibility with RPA correction is shown by the black dashed line.
({\bf d}) The susceptibility $\chi^{+-}_{\mathrm{AFM}}(\bm{q})$ as a function of the rotation angle $\theta$, which diverges logarithmically at the critical angle $\theta = 16^{\circ}$.
({\bf e}) The phase diagram in the $(\theta, U/t)$ plane where $U$ indicates the local Coulomb repulsion and $t$ indicates the nearest neighbor hopping amplitudes. Here the red (blue) dot denotes the critical point obtained by the RPA susceptibility (the self-consistent mean field study). At the critical angle the system show a magnetic instability even in the presence of an infinitesimally small interaction. 
  } \label{fig:phasediagram}
\end{figure*}

\begin{figure*}[t]
\centering
\includegraphics[width=160mm]{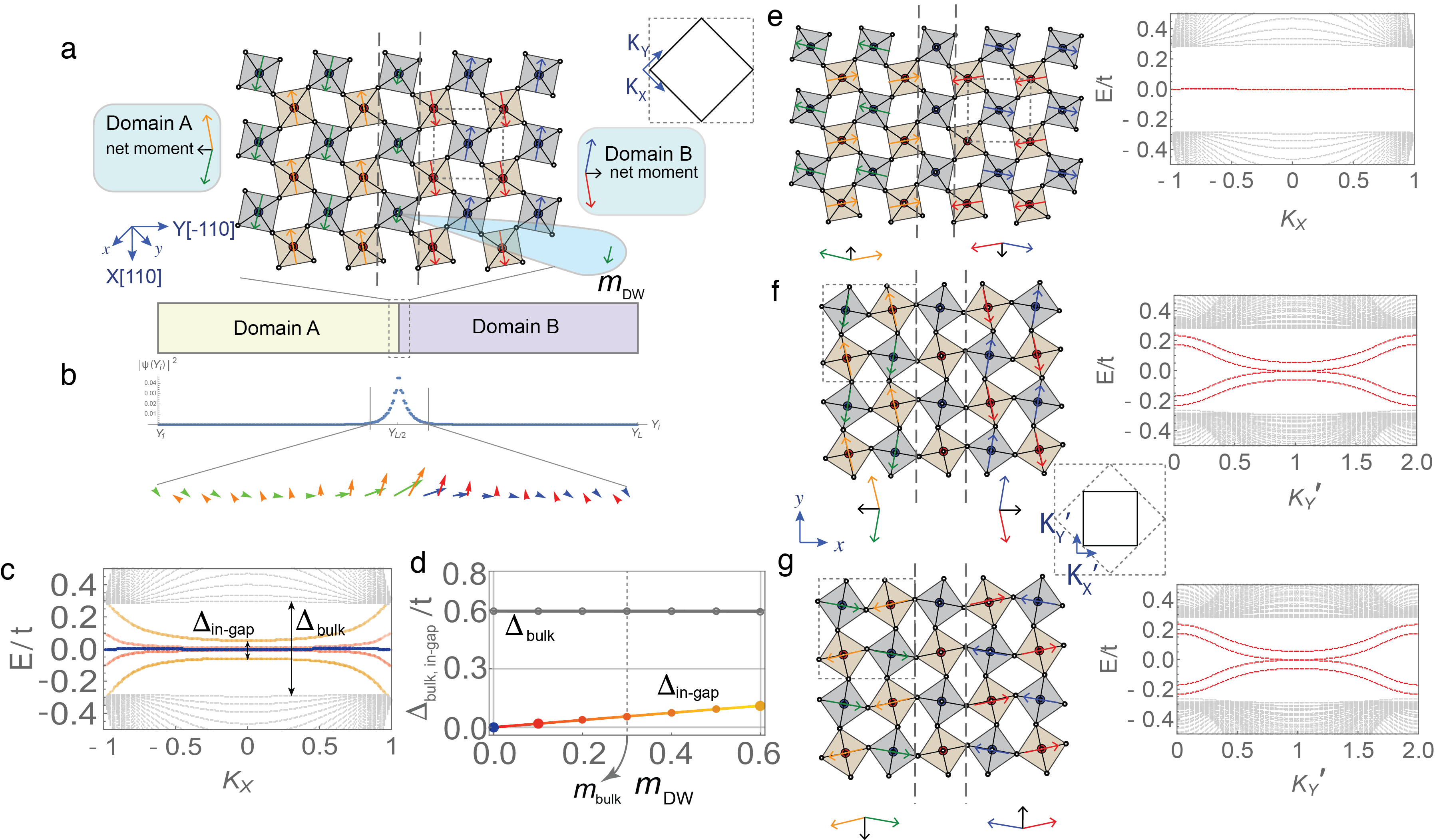}
\caption{
{\bf One dimensional (1D) soliton modes along magnetic domain walls (DWs).}
({\bf a}) A schematic figure showing a $1D$ DW parallel to the $X$ axis. The green and orange arrows indicate the magnetic moments in the domain $A$ with the net magnetic moment pointing perpendicular to the DW. The red and blue arrows are the magnetic moments in the domain B with the net magnetic moment opposite to that of the domain A.
({\bf b}) Spatial distribution (along the $Y$ direction) of the wave function amplitude of a soliton state localized at the DW.
({\bf c}) Energy spectrum of the system shown in ({\bf a}). The orange, red, blue lines show the dispersion of the in-gap states when the local magnetic moment at the DW, $m_{\textrm{DW}}$, is 0.6, 0.1, 0, respectively.
({\bf d}) The bulk gap ($\Delta_{\text{bulk}}$) v.s. the gap between in-gap states ($\Delta_{\text{in-gap}}$) as a function of the magnitude of $m_{\textrm{DW}}$.
({\bf e}) A DW configuration and the relevant energy spectrum when the net magnetic moment in each domain is parallel to the DW.
({\bf f, g}) A DW configuration and the relevant energy spectrum when the DW is parallel to the $x$ axis and the net magnetic moment is perpendicular to the DW ({\bf f}) and parallel to the DW ({\bf g}), respectively.
}\label{fig:DW}
\end{figure*}

\nsection
{Self-consistent mean-field calculations}
The divergent susceptibility due to dispersionless DLN indicates that the metallic state has an instability to a gapped phase which breaks the crystal symmetry leading to AFM state. The specific ordering pattern suggested by the susceptibility calculation is the $ab$-plane canted AFM rather than $c$-axis collinear AFM as shown in Fig.~\ref{fig:phasediagram}{\bf c}. To verify the magnetic ground state, we have performed the numerical analysis by means of self-consistent mean-field calculations. We allow the order parameters to describe any type of magnetic ordering patterns, thus we set ${\bf m}^A = (m^A_x, m^A_y, m^A_z)$ for sublattice A and ${\bf m}^B = (m^B_x, m^B_y, m^B_z)$ for sublattice B within a monolayer. The other order parameters in the remaining three layers are chosen by assuming the well-known ``up-down-down-up'' ordering pattern for net ferromagnetic moments~\cite{carter_theory_2013,upupdndn_jkim}. The chemical potential $\mu$ is determined iteratively to ensure the half-filling condition. The tolerance factor for the numerical iteration is fixed to $10^{-5}$ to ensure the convergence of the order parameters and chemical potential for given $(\theta, U)$ during several hundreds of iteration times. The resulting phase diagram is shown in Fig.~\ref{fig:phasediagram}{\bf e}. In the regime for the $ab$-plane canted AFM phase, due to the spin anisotropy originating from interlayer-coupling, the arbitrary initial value including the $c$-axis AFM converges into the $ab$-plane canted AFM as a final solution. We also have confirmed that the total energy of the $ab$-plane canted AFM is lower than that of the $c$-axis AFM. The critical interaction $U_c$ from self-consistent calculation agrees well with that from RPA-corrected susceptibility calculations. Convergence to ordered phase is tricky near critical rotational angle $\theta_c$ within our mean-field calculation scheme. More sophisticated numerical calculation may be needed to elaborate the results near the critical rotational angle. However, the overall tendency of the critical interaction $U_c$ as a function of the rotational angle $\theta$ is consistent with each other as shown in Fig.~\ref{fig:phasediagram}{\bf e}.

\nsection
{Localized line states}
We have introduced the diagonal line states in Eq.~\eqref{eq:linestates} providing us the basic building blocks to formulate the localized wave functions in the square lattice. Here, we will show that appropriate linear combinations of such localized diagonal line states with positive (negative) slope correspond to the degenerate eigenstates along the BZ boundary satisfying $k_x + k_y = \pi$ $(k_x - k_y = \pi)$. From the localized diagonal line states in Eq.~\eqref{eq:linestates}, using the Fourier transformation $c_{\sigma}^{\dagger}(\bm{r})=\frac{1}{\sqrt{4N^{2}}}\sum_{\bm{k}}e^{-i\bm{k}\cdot\bm{r}}c_{\bm{k},\sigma}^{\dagger}$ we find
\begin{align}
\left|\Psi\right\rangle_{A,\sigma}^{\alpha}(\phi)&=\frac{1}{\sqrt{N}}\sum_{m=1}^{N}e^{i 2m\phi}\left|\Psi\right\rangle_{\ell_{\alpha}=2m,\sigma}^{\alpha}
\nonumber\\
&=\frac{1}{\sqrt{N}}\sum_{m=1}^{N}\frac{1}{\sqrt{2N}}\sum_{\bm{r}\in\ell_{\alpha}=2m}\frac{1}{2N}\sum_{\bm{k}}e^{i2m\phi}(-1)^{r_{x}}e^{-i\bm{k}\cdot\bm{r}}c_{\bm{k},\sigma}^{\dagger}\left|0\right\rangle,
\end{align}
where $\bm{r}=(r_{x},r_{x}+2m-1)$ ($\bm{r}=(r_{x},-r_{x}+2m-1)$) for $\alpha=p$ ($\alpha=n$) with $r_{x}=1,2,...,2N$.
For a diagonal line with positive slope, we obtain
\begin{align}
\left|\Psi\right\rangle_{A,\sigma}^{p}(\phi) &=\frac{\sqrt{2}}{(2N)^{2}}\sum_{\bm{k}}\sum_{m=1}^{N}\sum_{r_{x}=1}^{2N}e^{i2m\phi}(-1)^{r_{x}}e^{-i\bm{k}\cdot(r_{x},r_{x}+2m-1)}c_{\bm{k},\sigma}^{\dagger}\left|0\right\rangle,
\nonumber\\
&=\frac{\sqrt{2}}{(2N)^{2}}\sum_{\bm{k}}\sum_{m=1}^{N}\sum_{r_{x}=1}^{2N}e^{i[r_{x}(\pi-k_{x}-k_{y})+2m(\phi-k_{y})+k_{y}]}c_{\bm{k},\sigma}^{\dagger}\left|0\right\rangle,
\end{align}
which, in the thermodynamic limit, becomes
\begin{align}
\left|\Psi\right\rangle_{A,\sigma}^{p}(\phi) &=\frac{1}{\sqrt{2}}\sum_{\bm{k}}\delta_{0,\pi-k_{x}-k_{y}}\delta_{0,\phi-k_{y}}e^{ik_{y}}c_{\bm{k},\sigma}^{\dagger}\left|0\right\rangle,
\nonumber\\
&=\frac{1}{\sqrt{2}}e^{i\phi}c_{(k_{x}=\pi-\phi,k_{y}=\phi),\sigma}^{\dagger}\left|0\right\rangle,
\end{align}
In this way, we obtain the Bloch state $\left|\Psi\right\rangle_{A,\sigma}^{p}(\phi)$ which is defined along BZ boundary satisfying $k_x + k_y = \pi$. The same property holds for $\left|\Psi\right\rangle_{\tau,\sigma}^{\alpha}(\phi)$ by changing sublattice $\tau$, pseudo-spin $\sigma$ and slope $\alpha$ indices with the wave number $\phi$ defined along the BZ boundary satisfying $k_x + k_y = \pi$ $(k_x - k_y = \pi)$ when it comes to positive (negative) slope.
Finally, the explicit form of the critical rotational angle from the condition $t_2 = 2 t_3$ is given by
\begin{equation}
\theta_c = {1\over 2} \tan^{-1} {\sqrt{
   5 V_{dd\delta} - 8 V_{dd\delta'} + 4 V_{dd\pi} - 16 V_{dd\pi'} +
    3 V_{dd\sigma}} \over \sqrt{2} \sqrt{-2 V_{dd\delta} + 5 V_{dd\delta'} - 4 V_{dd\pi} +
    4 V_{dd\pi'} + 3 V_{dd\sigma'} }}.
\end{equation}

%

%********************************************************************************************************************
%\bibliography{iridates}{}

%********************************************************************************************************************

{\small \subsection*{ACKNOWLEDGEMENTS}
J.-H.P., S.H.L. and C.H.K. were supported by the Institute for Basic Science in Korea (Grant No. IBS-R009-D1).
H.J. was supported by the Basic Science Research Program of the National Research Foundation of Korea (NRF) (Grant No. 2016R1D1A1B03933255, No. 2017M3D1A1040828).
B.-J.Y. was supported by the Institute for Basic Science in Korea (Grant No. IBS-R009-D1) and Basic Science Research Program through the National Research Foundation of Korea (NRF) (Grant No. 0426-20170012, No.0426-20180011), and  the POSCO Science Fellowship of POSCO TJ Park Foundation (No.0426-20180002). This work was supported in part by the U.S. Army Research Office under Grant Number W911NF-18-1-0137.
}

%********************************************************************************************************************

\newpage

%%%%%%%%%%%%%%%%%%%%%%%%%%%%%%%%%%%%%%%%%%%%%%%%%%%%%%%%%%%%%%%%%%%%%%%%%%
%%%%%%%%%%%%%%%%%%%%%%%%%%%%%%%%%%%%%%%%%%%%%%%%%%%%%%%%%%%%%%%%%%%%%%%%%%%

\newpage

%%%%%%%%%%%%%%%%%%%%%%%%%%%%%%%%%%%%%%%%%%%%%%%%%%%%%%%%%%%%%%%%%%%%%%%%%%%%
%%%%%%%%%%%%%%%%%%%%%%%%%%%%%%%%%%%%%%%%%%%%%%%%%%%%%%%%%%%%%%%%%%%%%%%%%%%%%%%%%

\newpage

%%%%%%%%%%%%%%%%%%%%%%%%%%%%%%%%%%%%%%%%%%%%%%%%%%%%%%%%%%%%%%%%%%%%%%%%%%%%%
%%%%%%%%%%%%%%%%%%%%%%%%%%%%%%%%%%%%%%%%%%%%%%%%%%%%%%%%%%%%%%%%%%%%%%%%%%%%%

\newpage

%%%%%%%%%%%%%%%%%%%%%%%%%%%%%%%%%%%%%%%%%%%%%%%%%%%%%%%%%%%%%%%%%%%%%%%%%%%%%
%%%%%%%%%%%%%%%%%%%%%%%%%%%%%%%%%%%%%%%%%%%%%%%%%%%%%%%%%%%%%%%%%%%%%%%%%%%%%

\newpage

%%%%%%%%%%%%%%%%%%%%%%%%%%%%%%%%%%%%%%%%%%%%%%%%%%%%%%%%%%%%%%%%%%%%%%%%%%%%%
%%%%%%%%%%%%%%%%%%%%%%%%%%%%%%%%%%%%%%%%%%%%%%%%%%%%%%%%%%%%%%%%%%%%%%%%%%%%%

\newpage

%%%%%%%%%%%%%%%%%%%%%%%%%%%%%%%%%%%%%%%%%%%%%%%%%%%%%%%%%%%%%%%%%%%%%%%%%%%%%
%%%%%%%%%%%%%%%%%%%%%%%%%%%%%%%%%%%%%%%%%%%%%%%%%%%%%%%%%%%%%%%%%%%%%%%%%%%%%

\clearpage
\widetext
\begin{center}

\noindent
\large{\textbf{Supporting Information: \\Two-dimensional Peierls instability via zone boundary Dirac line nodes in layered perovskite oxides}}\\

\normalsize{~\\
  Jin-Hong Park,$^{1,*}$ Seung Hun Lee,$^{1,2,*}$ Choong H. Kim,$^{1,2}$ Hosub Jin,$^{3,\dagger}$ and Bohm-Jung Yang $^{1,2,4,\ddagger}$}
\small{~\\
$^1$\textit{Center for Correlated Electron Systems,\\ Institute for
  Basic Science (IBS), Seoul 08826, Korea}\\
$^2$\textit{Department of Physics and Astronomy,\\ Seoul National
  University, Seoul 08826, Korea}\\
$^3$\textit{Department of Physics, Ulsan National Institute of
  Science and Technology (UNIST),\\ 50 UNIST, Ulsan 44919, South Korea}\\
$^4$\textit{Center for Theoretical Physics (CTP),\\ Seoul National
  University, Seoul 08826, Korea}}

\end{center}

%********************************************************************************************************************

\setcounter{equation}{0}
\setcounter{figure}{0}
\renewcommand{\thefigure}{S\arabic{figure}}
\renewcommand{\theequation}{S\arabic{equation}}
\renewcommand{\bibnumfmt}[1]{[#1]}

%********************************************************************************************************************

\newpage
\noindent{\bf Role of multiple nonsymmorphic symmetries}
\\

To describe the symmetry of Sr$_2$IrO$_4$, let us use the $\sqrt{2}\times\sqrt{2}$ unit cell from the beginning.
Namely, there are two atoms in a unit cell.
Then we define a unit translation along $x$ and $y$ directions
in a way that the first Brillouin zone is defined as $-\pi \leq k_{x}, k_{y}\leq \pi$.
Namely, here $x$ and $y$ coordinates corresponds to the conventional $a$ and $b$ coordinates in previous literatures.

\nsection
{Point group symmetries.} 
The point group symmetry of the system is generated by
inversion $P$, and two glide mirrors $G_{x}$ and $G_{y}$. Including time-reversal symmetry,
space-time coordinates transforms as
in the following way.
\begin{align}
T&:(x,y,t)\rightarrow (x,y,-t)\times i\sigma_{y},
\nonumber\\
P&:(x,y,t)\rightarrow (-x,-y,t),
\nonumber\\
G_{x}&:(x,y,t)\rightarrow (-x+\frac{1}{2},y+\frac{1}{2},t)\times i\sigma_{x},
\nonumber\\
G_{y}&:(x,y,t)\rightarrow (x+\frac{1}{2},-y+\frac{1}{2},t)\times i\sigma_{y},
\end{align}

Equivalently, one can use two two-fold screw rotations $S_{x}\equiv G_{x}P$ and $S_{y}\equiv G_{y}P$
instead of $G_{x,y}$. $S_{x,y}$ transforms space-time coordinate as
\begin{align}
S_{x}&:(x,y,t)\rightarrow (x+\frac{1}{2},-y+\frac{1}{2},t)\times i\sigma_{x},
\nonumber\\
S_{y}&:(x,y,t)\rightarrow (-x+\frac{1}{2},y+\frac{1}{2},t)\times i\sigma_{y},
\end{align}

% ********************************************************************************************************************

\nsection
{Dirac line nodes on the Brillouin zone boundary.}
Here we prove the symmetry protection of the Dirac line nodes on $k_{x}=\pi$ or $k_{y}=\pi$ lines.
First, let us consider $k_{x}=\pi$ line.
On this line, the systems is invariant under $PT$, $G_{x}$, $S_{y}$.
Thus every band on the $k_{x}=\pi$ line can be labelled either by $G_{x}$ eigenvalue or by $S_{y}$ eigenvalues.
The $G_{x}$ and $S_{y}$ eigenvalues can be simultaneously determined only if these two symmetries commute.
In any case, let us use $G_{x}$ eigenvalues to label bands on the $k_{x}=\pi$ line.
From $G_{x}^{2}=-e^{ik_{y}}$
we find that $G_{x}$ has two eigenvalues $n_{x,\pm}(k_{y})=\pm i e^{i\frac{1}{2}k_{y}}$.
Then one can define the $G_{x}$ eigenstates in the following way.
\begin{align}
G_{x}|n_{x,\pm}(k_{y})\rangle=n_{x,\pm}(k_{y})|n_{x,\pm}(k_{y})\rangle
\end{align}
which can be satisfied on the $k_{x}=0$ and $k_{x}=\pi$ lines.

To understand the band connection on the $k_{x}=\pi$ line, it is useful to examine the commutation
relation between $PT$, $G_{x}$, $S_{y}$.
Again, these symmetries transform the space time coordinates as
\begin{align}
PT&:(x,y,t)\rightarrow (-x,-y,-t)\times i\sigma_{y},
\nonumber\\
G_{x}&:(x,y,t)\rightarrow (-x+\frac{1}{2},y+\frac{1}{2},t)\times i\sigma_{x},
\nonumber\\
S_{y}&:(x,y,t)\rightarrow (-x+\frac{1}{2},y+\frac{1}{2},t)\times i\sigma_{y},
\end{align}
Their product transforms the space-time coordiates as
\begin{align}
PTG_{x}&:(x,y,t)\rightarrow (x-\frac{1}{2},-y-\frac{1}{2},-t)\times (i\sigma_{y})(-i\sigma_{x}),
\nonumber\\
G_{x}PT&:(x,y,t)\rightarrow (x+\frac{1}{2},-y+\frac{1}{2},-t)\times (i\sigma_{x})(i\sigma_{y}),
\nonumber\\
PTS_{y}&:(x,y,t)\rightarrow (x-\frac{1}{2},-y-\frac{1}{2},-t)\times (i\sigma_{y})(i\sigma_{y}),
\nonumber\\
S_{y}PT&:(x,y,t)\rightarrow (x+\frac{1}{2},-y+\frac{1}{2},-t)\times (i\sigma_{y})(i\sigma_{y}),
\nonumber\\
S_{y}G_{x}&:(x,y,t)\rightarrow (x,y+1,t)\times (i\sigma_{y})(i\sigma_{x}),
\nonumber\\
G_{x}S_{y}&:(x,y,t)\rightarrow (x,y+1,t)\times (i\sigma_{x})(i\sigma_{y}),
\end{align}
which gives rise to the following commutation relations
\begin{align}
PTG_{x}&=e^{ik_{x}-ik_{y}}G_{x}PT,
\nonumber\\
PTS_{y}&=e^{ik_{x}-ik_{y}}S_{y}PT,
\nonumber\\
S_{y}G_{x}&=-G_{x}S_{y},
\end{align}
This commutation relation is valid in the whole momentum space.

Now we again focus on the $k_{x}=\pi$ line on which we have
\begin{align}\label{eqn:commutation}
PTG_{x}&=-e^{-ik_{y}}G_{x}PT,
\nonumber\\
PTS_{y}&=-e^{-ik_{y}}S_{y}PT,
\nonumber\\
S_{y}G_{x}&=-G_{x}S_{y},
\end{align}
thus we see that $G_{x}$ and $S_{y}$ cannot be diagonalized simultaneously.

First, let us compare $G_{x}$ eigenvalues of $|n_{x,\pm}(k_{y})\rangle$ and $PT|n_{x,\pm}(k_{y})\rangle$.
From Eq.~(\ref{eqn:commutation}), we find
\begin{align}\label{eqn:PT and Gx}
G_{x}PT|n_{x,\pm}(k_{y})\rangle&=-e^{ik_{y}}PTG_{x}|n_{x,\pm}(k_{y})\rangle
\nonumber\\
&=-e^{ik_{y}}PT\left[\pm i e^{i\frac{1}{2}k_{y}}|n_{x,\pm}(k_{y})\rangle\right]
\nonumber\\
&=\pm i e^{i\frac{1}{2}k_{y}}\left[PT|n_{x,\pm}(k_{y})\rangle\right]
\nonumber\\
&=n_{x,\pm}(k_{y})\left[PT|n_{x,\pm}(k_{y})\rangle\right]
\end{align}
thus a Kramers pair $|n_{x,\pm}(k_{y})\rangle$ and $PT|n_{x,\pm}(k_{y})\rangle$ have the same $G_{x}$ eigenvalues.

Now we compare $G_{x}$ eigenvalues of $|n_{x,\pm}(k_{y})\rangle$ and $S_{y}|n_{x,\pm}(k_{y})\rangle$
From the anti-commuation relation between $S_{y}$ and $G_{x}$, it is obvious that
\begin{align}\label{eqn:Sx and Gx}
G_{x}S_{y}|n_{x,\pm}(k_{y})\rangle&=-[n_{x,\pm}(k_{y})]S_{y}|n_{x,\pm}(k_{y})\rangle
\nonumber\\
&=[n_{x,\mp}(k_{y})]S_{y}|n_{x,\pm}(k_{y})\rangle
\end{align}
thus $|n_{x,\pm}(k_{y})\rangle$ and $S_{y}|n_{x,\pm}(k_{y})\rangle$ have different $G_{x}$ eigenvalues.

Since the system is invariant under $PT$, $S_{y}$, $G_{x}$ on the $k_{x}=\pi$ line, the four states
$|n_{x,\pm}(k_{y})\rangle$ and $PT|n_{x,\pm}(k_{y})\rangle$
and $S_{y}|n_{x,\pm}(k_{y})\rangle$ and $PTS_{y}|n_{x,\pm}(k_{y})\rangle$ with $G_{x}$ eigenvalues
$n_{x,\pm}(k_{y})$, $n_{x,\pm}(k_{y})$, $n_{x,\mp}(k_{y})$, $n_{x,\mp}(k_{y})$, respectively,
are all degenerate with the same energy, thus there should be a Dirac line node with four-fold degeneracy
on the $k_{x}=\pi$ line.

One can perform similar analysis on the $k_{y}=\pi$ line, and thus the DLN spanning the full BZ boundary can be understoond.
\\

%********************************************************************************************************************

\newpage
\noindent{\bf Slater-Koster parameter method}
\\
%{Supplementary Note 2. Slater-Koster parameter method.} 
%
We use Slater-Koster parameters within same sublattice $\tau$ as follows:
\begin{align}
\langle d_{yz, 0, \tau} | H | d_{yz, i, \tau} \rangle &= V_{dd\pi} \cos^2 (\theta - \phi) + V_{dd\delta} \sin^2 (\theta - \phi),
\nonumber \\
\langle d_{zx, 0, \tau} | H | d_{zx, i, \tau} \rangle &= V_{dd\delta} \cos^2 (\theta - \phi) + V_{dd\pi} \sin^2 (\theta - \phi),
\nonumber \\
\langle d_{xy, 0, \tau} | H | d_{xy, i, \tau} \rangle &= V_{dd\pi} \cos^2 (2(\theta - \phi))
\nonumber \\
&+ V_{dd\sigma} \sin^2 (2(\theta - \phi)).
\end{align}
This Slater-Koster parameters depend on the rotation angle $\theta$ and relative displacement angle $\phi$ between two adjacent orbitals where $d_{\lambda,0}$-orbital locates at the origin and $d_{\lambda,i}$-orbital at $\v r_i = (x_i, y_i)$. Thus $\phi$ is defined by
\begin{equation}
(\cos \phi, \sin \phi) = (x_i, y_i)/\sqrt{x_i^2 + y_i^2}.
\end{equation}
Similarly, we have the Slater-Koster parameters between different sublattices $\tau$ and $\overline{\tau}$ which are given by
\begin{align}
& \langle d_{yz, 0, \tau}| H |d_{yz, i, \overline{\tau}} \rangle = \langle d_{yz, 0,  \overline{\tau}}| H |d_{yz, i,\tau} \rangle
\nonumber \\
&= V_{dd\pi} \cos(\theta - \phi) \cos (\theta + \phi) - V_{dd\delta} \sin(\theta-\phi)\sin(\theta+\phi),
\nonumber \\
& \langle d_{zx, 0, \tau}| H |d_{zx, i, \overline{\tau}} \rangle = \langle d_{zx, 0, \overline{\tau}}| H |d_{zx, i, \tau} \rangle
\nonumber \\
&= V_{dd\delta} \cos(\theta - \phi) \cos (\theta + \phi) - V_{dd\pi} \sin(\theta-\phi)\sin(\theta+\phi),
\nonumber \\
& \langle d_{xy, 0, \tau}| H |d_{xy, i, \overline{\tau}} \rangle = \langle d_{xy, 0, \overline{\tau}}| H |d_{xy, i, \tau} \rangle
\nonumber \\
& = V_{dd\pi} \cos(2(\theta - \phi)) \cos (2(\theta + \phi))
\nonumber \\
&~~~ - V_{dd\sigma} \sin(2(\theta-\phi))\sin(2(\theta+\phi)),
\end{align}
The collective rotation allows the hopping between different orbitals of $d_{yz}$ and $d_{zx}$ as follows:
\begin{align}
& \langle d_{yz, 0, \tau}| H |d_{zx, i, \overline{\tau}} \rangle = - \langle d_{yz, 0, \overline{\tau}}| H |d_{zx, i, \tau} \rangle
\nonumber \\
& = -V_{dd\pi} \cos(\theta - \phi) \sin (\theta + \phi) - V_{dd\delta} \cos(\theta+\phi)\sin(\theta-\phi),
\nonumber \\
& \langle d_{zx, 0, \tau}| H |d_{yz, i, \overline{\tau}} \rangle = - \langle d_{zx, 0, \overline{\tau}}| H |d_{yz, i, \tau} \rangle
\nonumber \\
& = V_{dd\pi} \sin(\theta - \phi) \cos(\theta + \phi) + V_{dd\delta} \cos(\theta-\phi)\sin(\theta+\phi).
\nonumber \\
\end{align}

\newpage
\noindent{\bf DFT band structure calculations}

\begin{figure}[htbp]
\begin{center}
    \includegraphics[width=110mm]{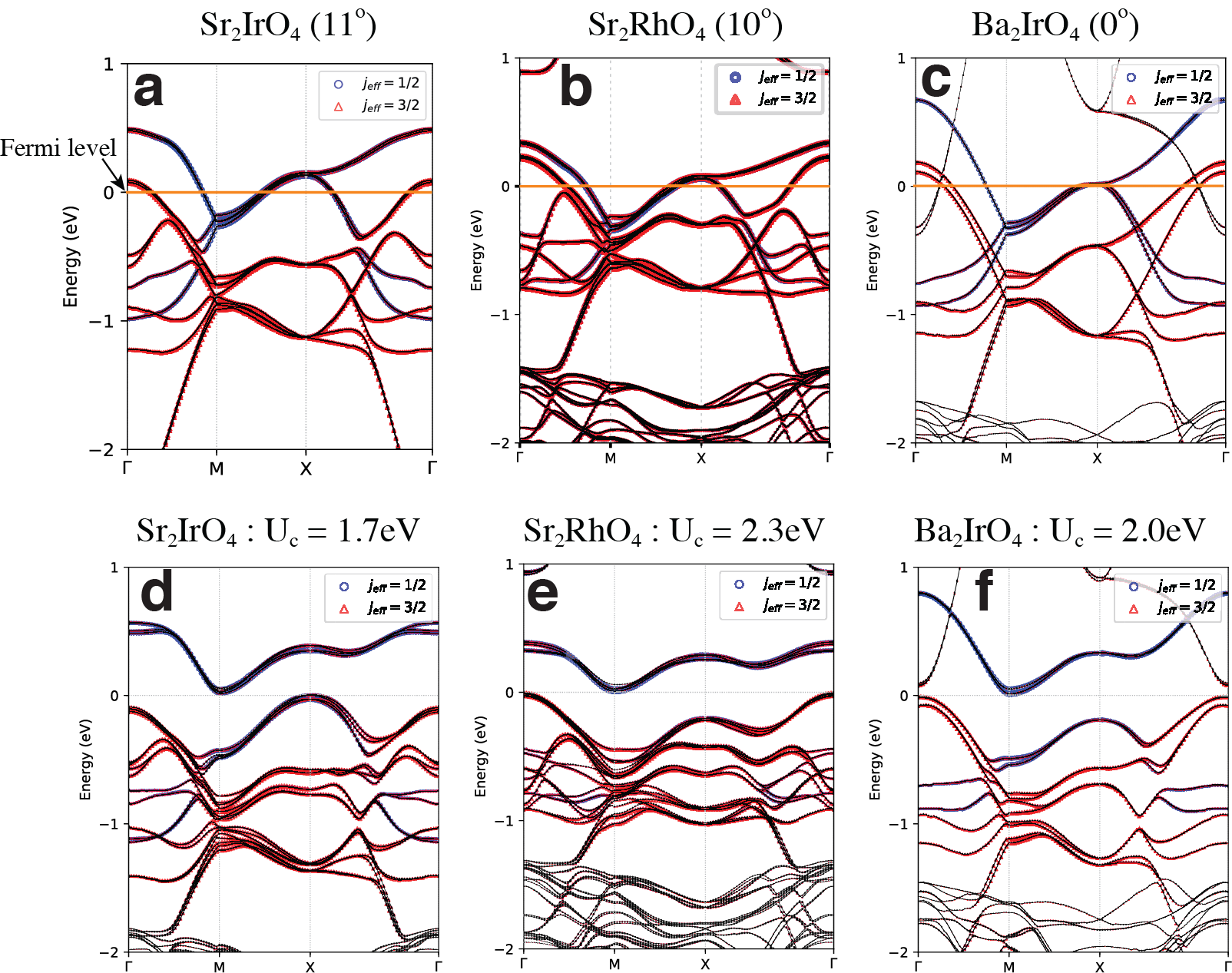}
    \caption{({\bf a}-{\bf c}) The DFT+SOC calculations for Sr$_2$IrO$_4$, Sr$_2$RhO$_4$, and Ba$_2$IrO$_4$ with rotational angles from experimental data. The orange line indicates the Fermi level. ({\bf d}-{\bf f}) The DFT+SOC+$U$ calculations for Sr$_2$IrO$_4$, Sr$_2$RhO$_4$, and Ba$_2$IrO$_4$ with rotational angles from experimental data. The critical $U$ for Sr$_2$IrO$_4$, Sr$_2$RhO$_4$, and Ba$_2$IrO$_4$ are $U_c = 1.7$eV, $2.3$eV, and $2.0$eV, respectively.}\label{fig:dft-soc}
\end{center}
\end{figure}

First, we provide the DFT calculations for Sr$_2$IrO$_4$, Sr$_2$RhO$_4$, and Ba$_2$IrO$_4$ with rotational angles from experimental data. The rotation angles for Sr$_2$IrO$_4$, Sr$_2$RhO$_4$, and Ba$_2$IrO$_4$ are $11^\circ$, $10^\circ$, and $0^\circ$, respectively. To compare the magnetic instability in these systems, one need to carefully take into account the Coulomb interaction $U$, and the influence of additional bands on the Fermi level together with the zone boundary Dirac line nodes. The paramagnetic band structures are determined by DFT+SOC calculations, and the critical $U$ for metal-insulator transition is obtained by DFT+SOC+$U$ calculations as shown in Fig.~\ref{fig:dft-soc}. 
One peculiar property of the Sr$_2$RhO$_4$ paramagnetic band structure compared to that of Sr$_2$IrO$_4$ is that the $J_{\textrm{eff}}=3/2$ states as well as the $J_{\textrm{eff}}=1/2$ states largely contribute to the Fermi surface due to the weak spin orbit coupling. Because of this, even if the zone boundary DLN becomes flat, its location is away from the Fermi level, which weakens the magnetic instability driven by DLN. This is also consistent with the fact that $U_c$ for Sr$_2$RhO$_4$ is bigger than that for Sr$_2$IrO$_4$.
On the other hand, in the case of Ba$_2$IrO$_4$, the system has an AFM ground state although there is no rotation distortion. In this case, one may expect that our theory based on the zone boundary DLN cannot be applied since the space group of the system remains symmorphic. However, even in this case, one can still use $\sqrt{2} \times \sqrt{2}$-type doubled unit cell to describe the magnetic instability since the property of the system is independent of the unit cell choice. In fact, according to the DFT+SOC calculations, the zone boundary Dirac line node is still present if the band structure is plotted by using the doubled unit cell as shown in Fig.~\ref{fig:dft-soc}{\bf c}. The $U_c$ for Ba$_2$IrO$_4$ is found to be bigger than that for Sr$_2$IrO$_4$, which is consistent with the fact that Sr$_2$IrO$_4$ has bigger rotation angle. If the rotation angle is artificially introduced, one can also observe the flattening of the zone boundary DLN in Ba$_2$IrO$_4$ as shown below.

% The bands of $J_{\textrm{eff}} = 3/2$ states are heavily participated near Fermi level due to the relatively weak spin-orbit interactions as shown in Fig.~\ref{fig:dft-soc}{\bf b}. The half-filled $J_{\textrm{eff}}=1/2$ bands is not perfectly satisfied, which is the main reason for  Sr$_2$RhO$_4$ to be a metal. Meanwhile, Ba$_2$IrO$_4$ has no rotational distortion. It is natural to expect that there is no unit cell doubling resulting in no Dirac line node. But according to the DFT+SOC calculations as in Fig.~\ref{fig:dft-soc}{\bf c}, the Dirac line node along BZ boundary is still present, which is consistent with the tight-binding calculation in Fig.~3{\bf b}. The strong Hubbard interaction $U$ is responsible for Ba$_2$IrO$_4$ to become AFM insulator even though there is finite energy dispersion along BZ boundary.  The DFT+SOC+$U$ calculations for Sr$_2$IrO$_4$, Sr$_2$RhO$_4$, and Ba$_2$IrO$_4$ with rotational angles from experimental data. The critical $U$ are determined by DFT+SOC+$U$ implementations. The consequence of rotational distortion for Sr$_2$IrO$_4$, Sr$_2$RhO$_4$, and Ba$_2$IrO$_4$ are as follows.

\newpage

\begin{figure}[htbp]
\begin{center}
    \includegraphics[width=160mm]{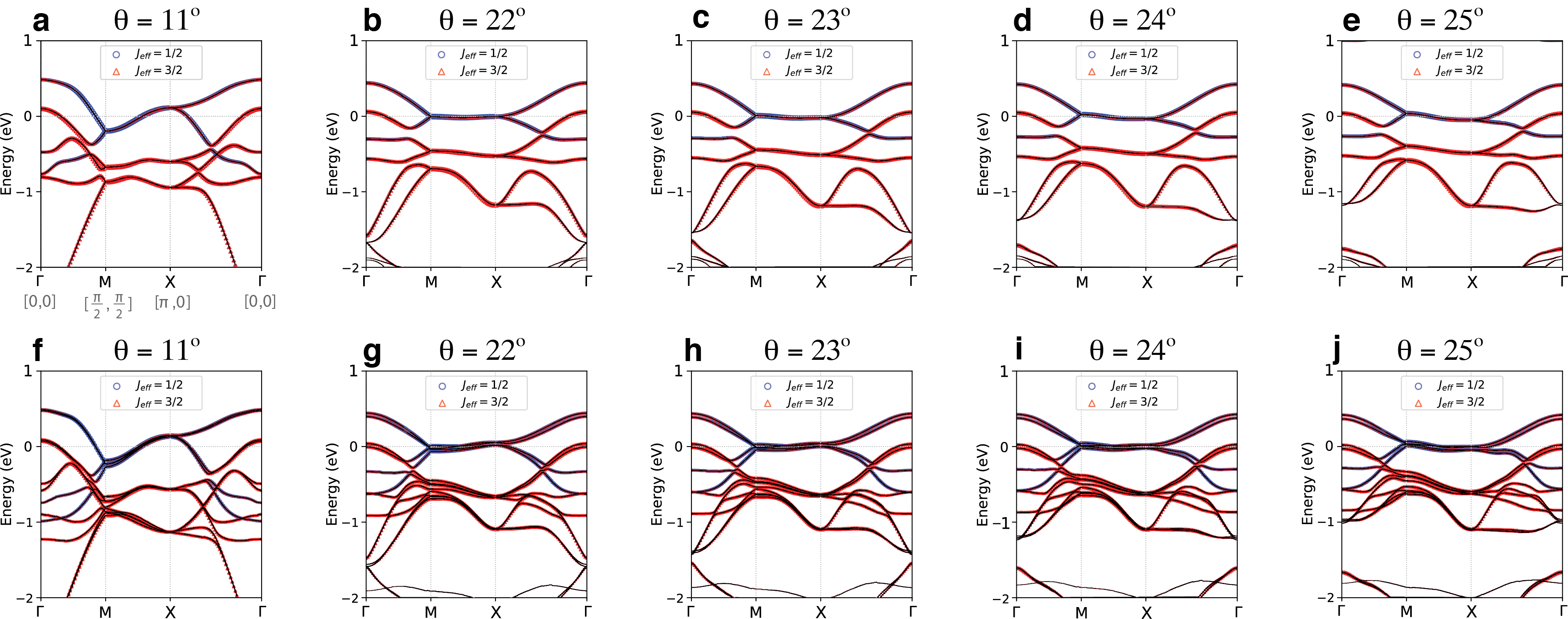}
    \caption{({\bf a-j}) DFT band structures of Sr$_2$IrO$_4$ as a function of in-plane rotation angle $\theta$, which is obtained by changing the Ir-O bond length while the in-plane lattice constant is fixed. ({\bf a-e}) Band structure of a monolayer. ({\bf f-j}) Band structure of the bulk. $J_{\textrm{eff}} =1/2$ (blue circle) and $J_{\textrm{eff}} =3/2$ (red triangle) bands are displayed by using different colors. The degenerate DLNs along the BZ boundary (M-X line) around the Fermi level become almost dispersionless at the critical angle $\theta\approx 23^\circ$ in both a monolayer and the bulk Sr$_2$IrO$_4$ as shown in Fig.~\ref{fig:dft-supp}{\bf c} and {\bf h}, respectively.}\label{fig:dft-supp}
\end{center}
\end{figure}

\newpage
\nsection{Sr$_2$IrO$_4$}
Here we provide additional DFT calculations varying the rotation angle $\theta$, which is obtained by changing the Ir-O bond length while fixing the in-plane lattice constant. The $\theta$-dependent evolution of DFT band structure for a monolayer is demonstrated in Fig.~\ref{fig:dft-supp}{\bf a-e}. During the evolution of $\theta$, the bandwidth of the DLNs changes consistent with Fig.~3{\bf h-l} as well as the tight-binding calculations in the main text. The emergence of flat DLNs is found in Fig.~\ref{fig:dft-supp}{\bf c}. The $\theta$-dependent evolution of DFT band structure for bulk Sr$_2$IrO$_4$ is demonstrated in Fig.~\ref{fig:dft-supp}{\bf f-j}, which agrees with Fig.~3{\bf m-q} in the main text. The emergence of almost flat bands from the DFT calculations strongly supports the robustness of our theory on the tunability of DLN via rotation distortion of octahedra in layered perovskite oxides.

\newpage

\begin{figure}[htbp]
\begin{center}
    \includegraphics[width=160mm]{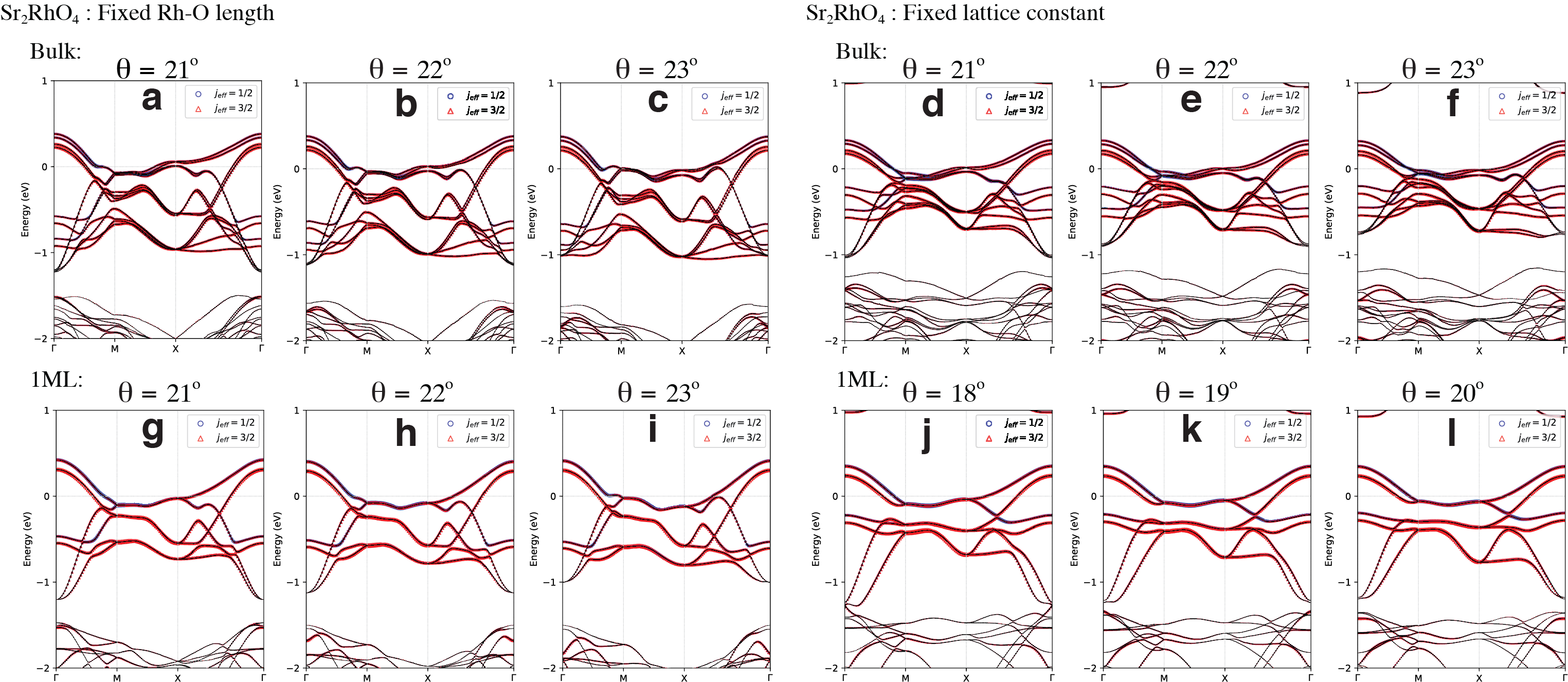}
    \caption{The evolution of the Sr$_2$RhO$_4$ band structure obtained by DFT calculations varying the rotation angle $\theta$. Here we implement two different ways of modifying the rotation angle both for bulk ({\bf a-f}) and for 1ML ({\bf g-l}).}\label{fig:dft-Rh214}
\end{center}
\end{figure}

\nsection{Sr$_2$RhO$_4$}
Here we provide DFT calculations of Sr$_2$RhO$_4$ varying the rotation angle $\theta$, which is obtained by changing in-plane lattice constant while the Rh-O bond length is fixed as in Figs.~\ref{fig:dft-Rh214}({\bf a-c} (bulk) and {\bf g-i} (1ML)), and by changing the Rh-O bond length while fixing the in-plane lattice constant in Figs.~\ref{fig:dft-Rh214}({\bf d-f} (bulk) and {\bf j-l}(1ML)). The large contribution of $J_{\textrm{eff}}=3/2$ states to the Fermi surface prevents Sr$_2$RhO$_4$ to fulfill the condition for half-filled $J_{\textrm{eff}}=1/2$ states but the $\theta$-dependent evolution of DFT band structure is successfully demonstrated.

\newpage

\begin{figure}[htbp]
\begin{center}
    \includegraphics[width=160mm]{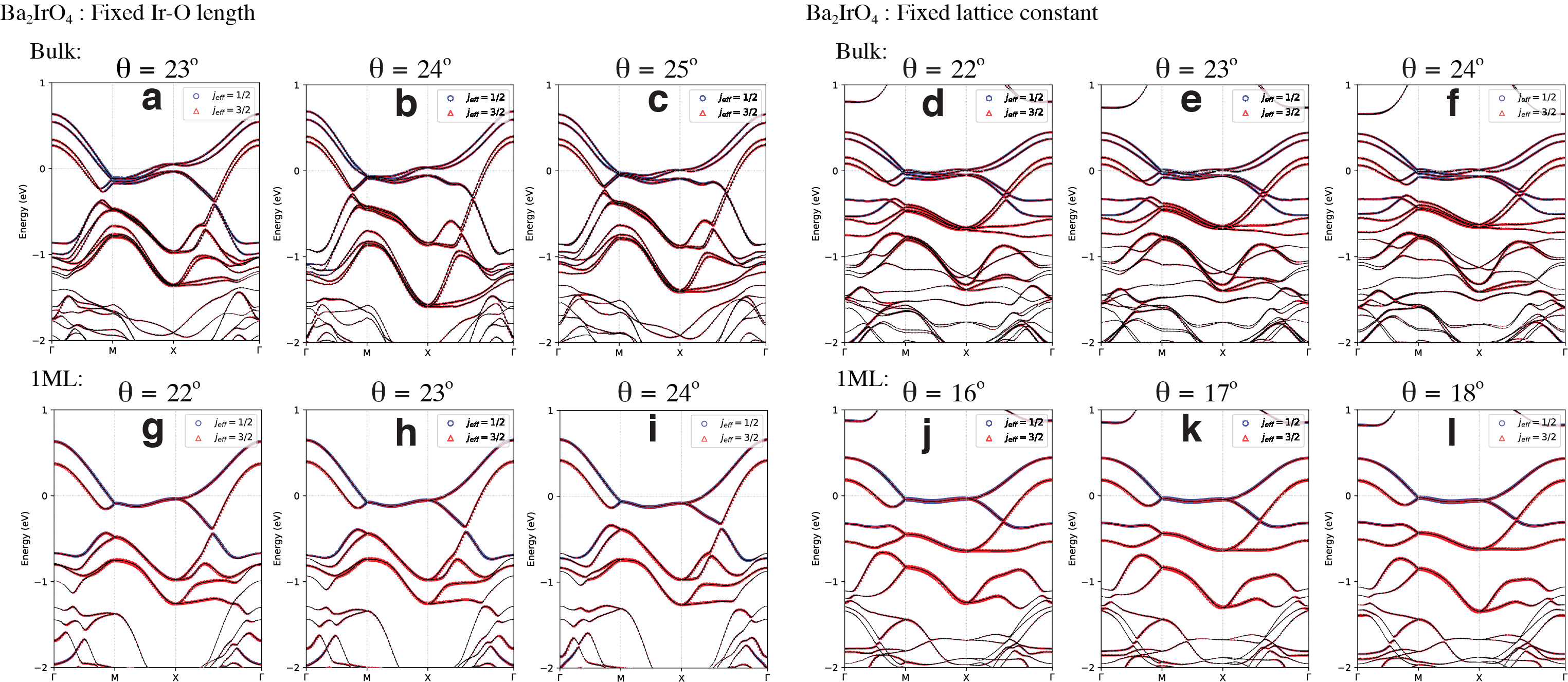}
    \caption{The evolution of the Ba$_2$IrO$_4$ band structure obtained by DFT calculations varying the rotation angle $\theta$. Here we implement two different ways of modifying the rotation angle both for bulk ({\bf a-f}) and for 1ML ({\bf g-l}).}\label{fig:dft-Ba214}
\end{center}
\end{figure}

\nsection{Ba$_2$IrO$_4$}
Here we provide DFT band structure calculations of Ba$_2$IrO$_4$ varying the rotation angle $\theta$, which is obtained by changing in-plane lattice constant while the Ir-O bond length is fixed as in Figs.~\ref{fig:dft-Rh214}({\bf a-c} (bulk) and {\bf g-i} (1ML)), and by changing the Ir-O bond length while fixing the in-plane lattice constant in Figs.~\ref{fig:dft-Rh214}({\bf d-f} (bulk) and {\bf j-l}(1ML)). One can find the almost flat band along BZ boundary for 1ML of $\theta = 16^\circ$ as shown in Fig.~\ref{fig:dft-Ba214}{\bf j}.

\newpage

\noindent{\bf DFT band structure calculations including staggered tetragonal distortion}

The second harmonic generation measurement indicates that Sr$_2$IrO$_4$ has $I4_1/a$ space group, which is also supported by neutron diffraction studies. In particular, the neutron diffraction data reported in Ref.~\cite{Feng_Neutron_2015} identified the staggered pattern for tetragonal distortion of oxygen octahedron. Namely, in both A and B sublattices, the oxygen octahedron is elongated along the $c$-direction, but the $c/a$ ratios in A and B sublattices are slightly different. Here $c/a$ ratio indicates the ratio of the out of plane Ir-O bond length over the in-plane Ir-O bond length. Such a staggered distortion of oxygen octahedron breaks the two glide mirror symmetries, which may affect the stability of the zone boundary Dirac line node.

However, if we compare the actual Ir-O bond lengths at the two Ir sublattice sites, (Ir$_1$ and Ir$_2$), one can expect that the staggered tetragonal distortion has very tiny effect on the electronic properties. More explicitly, according to the neutron diffraction data shown in Table I of \cite{Feng_Neutron_2015}, the out-of-plane and in-plane Ir-O distance in the unit of angstrom are given by $(2.056, 1.981)$ for Ir$_1$, and $(2.057, 1.979)$ for Ir$_2$. The corresponding $c/a$ ratios are $c/a=1.038$ for Ir$_1$ and $c/a=1.039$ for Ir$_2$, respectively. Namely, the relative change of Ir-O bond distances for two iridium sites is on the order of $10^{-3}$, which is sufficient to produce superlattice peak for structure analysis but is too tiny to affect the bulk electronic properties.

\begin{figure}[htbp]
\begin{center}
    \includegraphics[width=140mm]{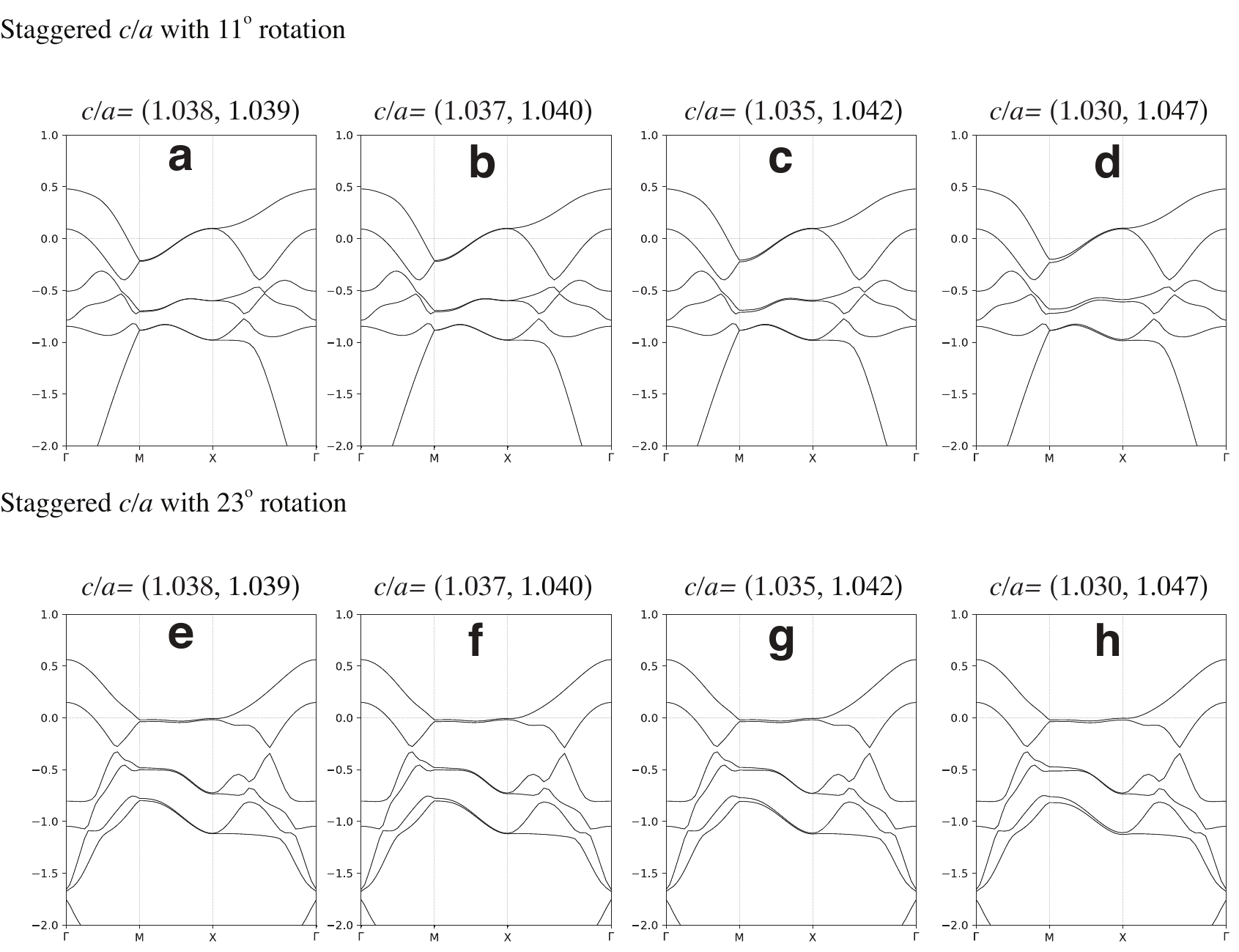}
    \caption{DFT+SOC calculations of Sr$_2$IrO$_4$ monolayer with staggered tetragonal distortions. ({\bf a}-{\bf d}) The band structure obtained by varying $c/a$ ratios with $\theta=11^\circ$. The energy splitting due to the staggered tetragonal distortion is negligible in the energy scale of Dirac line node dispersion. ({\bf e}-{\bf h}) The band structure obtained by varying $c/a$ ratios with $\theta=23^\circ$ when the zone boundary DLN is flat. The energy splitting due to the staggered tetragonal distortion is negligible.}\label{fig:dft-tetra}
\end{center}
\end{figure}

To demonstrate the negligible influence of the staggered tetragonal distortion on the electronic band structure, we performed additional DFT+SOC calculations taking into account the staggered tetragonal distortion in Fig.~\ref{fig:dft-tetra}. 
When $c/a=(1.038, 1.039)$ in two sublattices, which is the experimental value, one can see that staggered distortion indeed has negligible effect on the band degeneracy of the zone boundary Dirac line node (on the MX line) as shown in Fig.~\ref{fig:dft-tetra}{\bf a}. Only when the staggering of $c/a$ ratio is increased artificially up to $c/a=(1.030, 1047)$ shown in Fig.~\ref{fig:dft-tetra}{\bf d}, the weak splitting of the band degeneracy along the MX direction can be observed. This clearly shows that the influence of the staggered tetragonal distortion on the electronic band structure is negligible and thus our theory of 2D Peierls instability can be applied to Sr$_2$IrO$_4$.

\newpage
\noindent{\bf Hamiltonian with interlayer hopping}
\\
The Hamiltonian with interlayer hopping reads
\begin{equation}
H^{ll'}_{\v k, \theta} = \bpm A_{\v k, \theta} & B_{\v k, \theta} & 0 & e^{i k_z c} C_{\v k, \theta}^\dag \\
B_{\v k, \theta}^\dag & A_{\v k, \theta} & C_{\v k, \theta} & 0 \\
0 & C_{\v k, \theta}^\dag &  A_{\v k, \theta} & B_{\v k, \theta} \\
e^{-i k_z c} C_{\v k, \theta} & 0 & B_{\v k, \theta}^\dag &  A_{\v k, \theta} \\
\epm, \label{eq:interlayer-H}
\end{equation}
where
\begin{equation}
  \begin{split}
A_{\v k, \theta} &= \varepsilon_{\v k, \theta}^a \tau^0 \sigma^0 +
\varepsilon_{\v k, \theta}^{ad} \tau^x \sigma^0 + \varepsilon_{\v k, \theta}^{ad'}
\tau^y \sigma^z,
\\
B_{\v k, \theta} &= \varepsilon_{\v k, \theta}^b \tau^0 \sigma^0 +
\varepsilon_{\v k, \theta}^{bd} \tau^x \sigma^0 + \varepsilon_{\v k, \theta}^{bz}
\tau^y \sigma^z
\\
& + \varepsilon_{\v k, \theta}^{by} \tau^y \sigma^y +
\varepsilon_{\v k, \theta}^{bx} \tau^y \sigma^x,
\\
B_{\v k, \theta}^\dag &= \varepsilon_{\v k, \theta}^b \tau^0 \sigma^0 +
\varepsilon_{\v k, \theta}^{bd} \tau^x \sigma^0 + \varepsilon_{\v k, \theta}^{bz}
\tau^y \sigma^z
\\
&- \varepsilon_{\v k, \theta}^{by} \tau^y \sigma^y -
\varepsilon_{\v k, \theta}^{bx} \tau^y \sigma^x,
\\
C_{\v k, \theta} &= \varepsilon_{\v k, \theta}^c \tau^0 \sigma^0 +
\varepsilon_{\v k, \theta}^{cd} \tau^x \sigma^0 + \varepsilon_{\v k, \theta}^{cz}
\tau^y \sigma^z
\\
&+ \varepsilon_{\v k, \theta}^{cy} \tau^y \sigma^y +
\varepsilon_{\v k, \theta}^{cx} \tau^y \sigma^x,
\\
C_{\v k, \theta}^\dag &= \varepsilon_{\v k, \theta}^c \tau^0 \sigma^0 +
\varepsilon_{\v k, \theta}^{cd} \tau^x \sigma^0 + \varepsilon_{\v k, \theta}^{cz}
\tau^y \sigma^z
\\
&- \varepsilon_{\v k, \theta}^{cy} \tau^y \sigma^y -
\varepsilon_{\v k, \theta}^{cx} \tau^y \sigma^x.\\
\end{split}
\end{equation}
Here, $\tau^i$ and $\sigma^i$ with $i = 0,x,y,z$ indicate the Pauli
matrices acting on the
sublattice spaces and $J_{eff} =1/2$ spaces, respectively.
To fully express Eq.~\eqref{eq:interlayer-H} with Pauli matrices,
$\rho^i$ and $\eta^i$ is introduced without any physical meaning:
\begin{equation}
  \begin{split}
& \rho^0 \eta^0 =
\bpm
1 & 0&0&0 \\
0& 1&0&0 \\
0&0&1&0 \\
0&0&0&1\\
\epm,
~ \rho^x \eta^0 =
\bpm
0 & 1&0&0 \\
1& 0&0&0 \\
0&0&0&1 \\
0&0&1&0\\
\epm,
\\
& (i\rho^y) \eta^0 =
\bpm 0 & 1&0&0 \\
-1& 0&0&0 \\
0&0&0&1 \\
0&0&-1&0\\
\epm, 
~ \rho^x \eta^x =
\bpm
0 & 0&0&1 \\
0& 0&1&0 \\
0&1&0&0 \\
1&0&0&0 \\
\epm,
\\
&\rho^y \eta^y =
\bpm 0 & 0&0&-1 \\
0& 0&1&0 \\
0&1&0&0 \\
-1&0&0&0\\
\epm.
\end{split}
\end{equation}
With this, we can write down Eq.~\eqref{eq:interlayer-H} as a compact
form
\begin{equation}
  \begin{split}
H^{ll'}_{\v k, \theta}
&= \varepsilon_{\v k, \theta}^a \tau^0 \sigma^0 \rho^0
\eta^0 + \varepsilon_{\v k, \theta}^{ad} \tau^x \sigma^0 \rho^0
\eta^0 + \varepsilon_{\v k, \theta}^{ad'}
\tau^y \sigma^z \rho^0
\eta^0
\\
&+ \varepsilon_{\v k, \theta}^b \tau^0 \sigma^0 \rho^x
\eta^0 +
\varepsilon_{\v k, \theta}^{bd} \tau^x \sigma^0 \rho^x
\eta^0 + \varepsilon_{\v k, \theta}^{bz}
\tau^y \sigma^z \rho^x
\eta^0
\\
&+ \varepsilon_{\v k, \theta}^{by} \tau^y \sigma^y (i\rho^y)
\eta^0+
\varepsilon_{\v k, \theta}^{bx} \tau^y \sigma^x (i\rho^y)
\eta^0
\\
&+ \varepsilon_{\v k, \theta}^c \tau^0 \sigma^0 \rho^x
\eta^x +
\varepsilon_{\v k, \theta}^{cd} \tau^x \sigma^0 \rho^x
\eta^x + \varepsilon_{\v k, \theta}^{cz}
\tau^y \sigma^z \rho^x
\eta^x
\\
&+ \varepsilon_{\v k, \theta}^{cy} \tau^y \sigma^y \rho^y
\eta^y+
\varepsilon_{\v k, \theta}^{cx} \tau^y \sigma^x \rho^y
\eta^y.
\end{split}
\end{equation}
We are now in position to construct the Green's function and take the
trace of matrix for spin susceptibility. \\

\newpage
\noindent{\bf Susceptibility}
\\
The general form of susceptibility depends on the sublattice, spin, and
layer indices:
\begin{equation}
\chi^{ij}_{\alpha \alpha',l l'} (\v q) = -\int_{0}^\beta d \tau
\langle S^i_{\alpha l} (\v q,
\tau) S^j_{\alpha' l'} (- \v q, 0) \rangle,
\end{equation}
where $\alpha, \alpha'$ and $l, l'$ indicate the sublattice and layer
indices, respectively. The relevant physical susceptibility can be
expressed as

\begin{equation}
  \begin{split}
\chi^{zz}_{\mathrm{AFM}} (\v q, i \nu_n) =&
{1 \over \beta V} \sum_{\v k} \sum_{i \omega} \mathrm{Tr} [G(\v k, i
\omega_n) (\tau^z \sigma^z \rho^z \eta^z)
\\
& \times G(\v k + \v q, i
\omega_n + i \nu_n) (\tau^z \sigma^z
\rho^z \eta^z)],
\\
\chi^{zz}_{\mathrm{FM}} (\v q, i \nu_n) =&
{1 \over \beta V} \sum_{\v k} \sum_{i \omega} \mathrm{Tr} [G(\v k, i
\omega_n) (\tau^0 \sigma^z \rho^z \eta^z)
\\
& \times G(\v k + \v q, i
\omega_n + i \nu_n) (\tau^0 \sigma^z
\rho^z \eta^z)],
\\
\chi^{+-}_{\mathrm{AFM}} (\v q, i \nu_n) =&
{1 \over \beta V} \sum_{\v k} \sum_{i \omega} \mathrm{Tr} [G(\v k, i
\omega_n) (\tau^z (\sigma^x + i \sigma^y) \rho^z \eta^z)
\\
& \times G(\v k + \v q, i
\omega_n + i \nu_n) (\tau^z (\sigma^x - i \sigma^y) \rho^z \eta^z)],
\\
\chi^{+-}_{\mathrm{FM}} (\v q, i \nu_n) =&
{1 \over \beta V} \sum_{\v k} \sum_{i \omega} \mathrm{Tr} [G(\v k, i
\omega_n) (\tau^0 (\sigma^x + i \sigma^y) \rho^z \eta^z)
\\
& \times G(\v k + \v q, i
\omega_n + i \nu_n) (\tau^0 (\sigma^x - i \sigma^y ) \rho^z \eta^z)].
\label{eq:sus}
\end{split}
\end{equation}
With antiferromagnetic  interlayer coupling fixed to $\rho^z \eta^z = \bpm1&0&0&0 \\ 0&-1&0&0 \\ 0&0&-1&0 \\
0&0&0&1\\ \epm $ in a manner of the up-up-down-down spin
configuration, we focus on spin susceptibility in 2D. In that sense,
the four susceptibility of $\chi^{zz}_{\mathrm{AFM}}, \chi^{zz}_{\mathrm{FM}},
\chi^{+-}_{\mathrm{AFM}}, \chi^{+-}_{\mathrm{FM}}  $ are taken into account.
%This is justified to the fact that two-dimensional magnetic coupling is an order of magnitude larger
%than interlayer coupling~\cite{takayama2016}. The dominant contribution
%for magnetic ordering is within single layer so AFM and FM is
%meaningful in 2D. For interlayer, we assume the up-down-down-up spin configuration
%along $c$-axis by setting in Eq.~\eqref{eq:sus}. 
Further evaluation of Eq.~\eqref{eq:sus} reads
\begin{equation}
  \begin{split}
\mathrm{Tr}[\cdots]_{\mathrm{AFM}}^{zz} &= 16\Bigl((i \omega_n -
\varepsilon^0_{\v k}) (i \omega_n +i \nu_n - \varepsilon^0_{\v k +\v
  q}) +\Delta^1_{\v k, \v q} \Bigr)
\\
& \times (N_{\v k} (i \omega_n) N_{\v k + \v q} (i
\omega_n +i \nu_n))^{-1},
\\
\mathrm{Tr}[\cdots]_{\mathrm{FM}}^{zz} &= 16\Bigl((i \omega_n -
\varepsilon^0_{\v k}) (i \omega_n +i \nu_n - \varepsilon^0_{\v k +\v
  q}) + \Delta^2_{\v k, \v q} \Bigr)
\\
& \times (N_{\v k} (i \omega_n) N_{\v k + \v q} (i
\omega_n +i \nu_n))^{-1},
\\
\mathrm{Tr}[\cdots]_{\mathrm{AFM}}^{+-} &= 32\Bigl((i \omega_n -
\varepsilon^0_{\v k}) (i \omega_n +i \nu_n - \varepsilon^0_{\v k +\v
  q}) + \Delta^3_{\v k, \v q} \Bigr)
\\
&\times (N_{\v k} (i \omega_n) N_{\v k + \v q} (i
\omega_n +i \nu_n))^{-1},
\\
\mathrm{Tr}[\cdots]_{\mathrm{FM}}^{+-} &= 32\Bigl((i \omega_n -
\varepsilon^0_{\v k}) (i \omega_n +i \nu_n - \varepsilon^0_{\v k +\v
  q}) + \Delta^4_{\v k, \v q}  \Bigr)
\\
& \times (N_{\v k} (i \omega_n) N_{\v k + \v q} (i
\omega_n +i \nu_n))^{-1},
\end{split}
\end{equation}
where we have
\begin{equation}
  \begin{split}
\Delta^1_{\v k, \v q} &=  -
\varepsilon_{\v k}^{\mathrm{ad}} \varepsilon_{\v k + \v
  q}^{\mathrm{ad}} - \varepsilon_{\v k}^{\mathrm{ad'}} \varepsilon_{\v k + \v
  q}^{\mathrm{ad'}} - \varepsilon_{\v k}^{\mathrm{b}} \varepsilon_{\v k + \v
  q}^{\mathrm{b}} + \varepsilon_{\v k}^{\mathrm{bd}} \varepsilon_{\v k + \v
  q}^{\mathrm{bd}}
\\
&+ \varepsilon_{\v k}^{\mathrm{bx}} \varepsilon_{\v k + \v
  q}^{\mathrm{bx}}  + \varepsilon_{\v k}^{\mathrm{by}} \varepsilon_{\v k + \v
  q}^{\mathrm{by}} + \varepsilon_{\v k}^{\mathrm{bz}} \varepsilon_{\v k + \v
  q}^{\mathrm{bz}} + \varepsilon_{\v k}^{\mathrm{c}} \varepsilon_{\v k + \v
  q}^{\mathrm{c}}
\\
& -  \varepsilon_{\v k}^{\mathrm{cd}} \varepsilon_{\v k + \v
  q}^{\mathrm{cd}} +  \varepsilon_{\v k}^{\mathrm{cx}} \varepsilon_{\v k + \v
  q}^{\mathrm{cx}} +  \varepsilon_{\v k}^{\mathrm{cy}} \varepsilon_{\v k + \v
  q}^{\mathrm{cy}}  -  \varepsilon_{\v k}^{\mathrm{cz}} \varepsilon_{\v k + \v
  q}^{\mathrm{cz}},
\\
\Delta^2_{\v k, \v q} &=
\varepsilon_{\v k}^{\mathrm{ad}} \varepsilon_{\v k + \v
  q}^{\mathrm{ad}} + \varepsilon_{\v k}^{\mathrm{ad'}} \varepsilon_{\v k + \v
  q}^{\mathrm{ad'}} - \varepsilon_{\v k}^{\mathrm{b}} \varepsilon_{\v k + \v
  q}^{\mathrm{b}} - \varepsilon_{\v k}^{\mathrm{bd}} \varepsilon_{\v k + \v
  q}^{\mathrm{bd}}
\\
&- \varepsilon_{\v k}^{\mathrm{bx}} \varepsilon_{\v k + \v
  q}^{\mathrm{bx}}  - \varepsilon_{\v k}^{\mathrm{by}} \varepsilon_{\v k + \v
  q}^{\mathrm{by}} - \varepsilon_{\v k}^{\mathrm{bz}} \varepsilon_{\v k + \v
  q}^{\mathrm{bz}} + \varepsilon_{\v k}^{\mathrm{c}} \varepsilon_{\v k + \v
  q}^{\mathrm{c}}
\\
&+  \varepsilon_{\v k}^{\mathrm{cd}} \varepsilon_{\v k + \v
  q}^{\mathrm{cd}} -  \varepsilon_{\v k}^{\mathrm{cx}} \varepsilon_{\v k + \v
  q}^{\mathrm{cx}} -  \varepsilon_{\v k}^{\mathrm{cy}} \varepsilon_{\v k + \v
  q}^{\mathrm{cy}} +  \varepsilon_{\v k}^{\mathrm{cz}} \varepsilon_{\v k + \v
  q}^{\mathrm{cz}},
\\
\Delta^3_{\v k, \v q} &= -
\varepsilon_{\v k}^{\mathrm{ad}} \varepsilon_{\v k + \v
  q}^{\mathrm{ad}} + \varepsilon_{\v k}^{\mathrm{ad'}} \varepsilon_{\v k + \v
  q}^{\mathrm{ad'}} - \varepsilon_{\v k}^{\mathrm{b}} \varepsilon_{\v k + \v
  q}^{\mathrm{b}} + \varepsilon_{\v k}^{\mathrm{bd}} \varepsilon_{\v k + \v
  q}^{\mathrm{bd}}
\\
&- \varepsilon_{\v k}^{\mathrm{bz}} \varepsilon_{\v k + \v
  q}^{\mathrm{bz}} + \varepsilon_{\v k}^{\mathrm{c}} \varepsilon_{\v k + \v
  q}^{\mathrm{c}} -  \varepsilon_{\v k}^{\mathrm{cd}} \varepsilon_{\v k + \v
  q}^{\mathrm{cd}} +  \varepsilon_{\v k}^{\mathrm{cz}} \varepsilon_{\v k + \v
  q}^{\mathrm{cz}},
\\
\Delta^4_{\v k, \v q} &=
\varepsilon_{\v k}^{\mathrm{ad}} \varepsilon_{\v k + \v
  q}^{\mathrm{ad}} - \varepsilon_{\v k}^{\mathrm{ad'}} \varepsilon_{\v k + \v
  q}^{\mathrm{ad'}} - \varepsilon_{\v k}^{\mathrm{b}} \varepsilon_{\v k + \v
  q}^{\mathrm{b}} - \varepsilon_{\v k}^{\mathrm{bd}} \varepsilon_{\v k + \v
  q}^{\mathrm{bd}}
\\
&+ \varepsilon_{\v k}^{\mathrm{bz}} \varepsilon_{\v k + \v
  q}^{\mathrm{bz}} + \varepsilon_{\v k}^{\mathrm{c}} \varepsilon_{\v k + \v
  q}^{\mathrm{c}} +  \varepsilon_{\v k}^{\mathrm{cd}} \varepsilon_{\v k + \v
  q}^{\mathrm{cd}}-  \varepsilon_{\v k}^{\mathrm{cz}} \varepsilon_{\v k + \v
  q}^{\mathrm{cz}},
\\
N_{\v k} (i \omega_n) &= (i \omega_n - \varepsilon^0_{\v k})^2 - \Bigl(
\varepsilon_{\v k}^{\mathrm{ad}} \Bigr)^2 - \Bigl(
\varepsilon_{\v k}^{\mathrm{ad'}} \Bigr)^2 - \Bigl(
\varepsilon_{\v k}^{\mathrm{b}} \Bigr)^2
\\
& - \Bigl(
\varepsilon_{\v k}^{\mathrm{bd}} \Bigr)^2 - \Bigl(
\varepsilon_{\v k}^{\mathrm{bx}} \Bigr)^2 - \Bigl(
\varepsilon_{\v k}^{\mathrm{by}} \Bigr)^2 - \Bigl(
\varepsilon_{\v k}^{\mathrm{bz}} \Bigr)^2
\\
&- 
\Bigl(
\varepsilon_{\v k}^{\mathrm{c}} \Bigr)^2 - \Bigl(
\varepsilon_{\v k}^{\mathrm{cd}} \Bigr)^2 - \Bigl(
\varepsilon_{\v k}^{\mathrm{cx}} \Bigr)^2 - \Bigl(
\varepsilon_{\v k}^{\mathrm{cy}} \Bigr)^2 - \Bigl(
\varepsilon_{\v k}^{\mathrm{cz}} \Bigr)^2
\\
&=  \Bigl( (i \omega_n - \varepsilon^0_{\v k}) + Z_{\v k} \Bigr) \Bigl( (i
\omega_n - \varepsilon^0_{\v k}) - Z_{\v k} \Bigr), 
\label{eq:den}
\end{split}
\end{equation}
with
\begin{equation}
  \begin{split}
Z_{\v k} &= \Bigl[ \Bigl(
\varepsilon_{\v k}^{\mathrm{ad}} \Bigr)^2 + \Bigl(
\varepsilon_{\v k}^{\mathrm{ad'}} \Bigr)^2 + \Bigl(
\varepsilon_{\v k}^{\mathrm{b}} \Bigr)^2 + \Bigl(
\varepsilon_{\v k}^{\mathrm{bd}} \Bigr)^2 + \Bigl(
\varepsilon_{\v k}^{\mathrm{bx}} \Bigr)^2
\\
&+ \Bigl(
\varepsilon_{\v k}^{\mathrm{by}} \Bigr)^2+ \Bigl(
\varepsilon_{\v k}^{\mathrm{bz}} \Bigr)^2 + \Bigl(
\varepsilon_{\v k}^{\mathrm{c}} \Bigr)^2 + \Bigl(
\varepsilon_{\v k}^{\mathrm{cd}} \Bigr)^2 + \Bigl(
\varepsilon_{\v k}^{\mathrm{cx}} \Bigr)^2
\\
&+ \Bigl(
\varepsilon_{\v k}^{\mathrm{cy}} \Bigr)^2 + \Bigl(
\varepsilon_{\v k}^{\mathrm{cz}} \Bigr)^2 \Bigr]^{1/2}.
\label{eq:z}
\end{split}
\end{equation}
The replacement $\v k \rightarrow \v k + \v q$ and $i \omega_n \rightarrow
i \omega_n +i \nu_n$ in Eq.~\eqref{eq:den} and \eqref{eq:z} can yield
$N_{\v k + \v q} (i \omega_n + i \nu_n)$ and $Z_{\v k + \v q}$.
After Matsubara summation, we arrive at
\begin{widetext}
  \begin{equation}
    \begin{split}
\chi^{zz}_{\mathrm{AFM}} (\v q, i \nu_n) &=
{1 \over V} \sum_{\v k} \Bigl[ {1
  \over Z^2_{\v k} - Z^4_{\v k + \v q} } \Bigl(4+ {4\Delta^1_{\v k, \v q}
  \over Z_{\v k} Z_{\v k + \v q}} \Bigr)  \Bigl( {1 \over e^{\beta
    Z^2_{\v k}} + 1 } - {1 \over e^{\beta
    Z^4_{\v k + \v q}} + 1 } \Bigr)
\\
& + {1 \over Z^1_{\v k} - Z^3_{\v k + \v q} } \Bigl(4+ {4\Delta^1_{\v k, \v q}
  \over Z_{\v k} Z_{\v k + \v q}} \Bigr)  \Bigl( {1 \over e^{\beta
    Z^1_{\v k}} + 1 } - {1 \over e^{\beta
    Z^3_{\v k + \v q}} + 1 } \Bigr)
\\
& + {1 \over Z^2_{\v k} - Z^3_{\v k + \v q} } \Bigl(4 - {4\Delta^1_{\v k, \v q}
  \over Z_{\v k} Z_{\v k + \v q}} \Bigr)  \Bigl( {1 \over e^{\beta
    Z^2_{\v k}} + 1 } - {1 \over e^{\beta
    Z^3_{\v k + \v q}} + 1 } \Bigr)
\\
& + {1 \over Z^1_{\v k} - Z^4_{\v k + \v q} } \Bigl(4 - {4\Delta^1_{\v k, \v q}
  \over Z_{\v k} Z_{\v k + \v q}} \Bigr)  \Bigl( {1 \over e^{\beta
    Z^1_{\v k}} + 1 } - {1 \over e^{\beta
    Z^4_{\v k + \v q}} + 1 } \Bigr)
\Bigr].
\end{split}
\label{eqS:sus-final}
\end{equation}
\end{widetext}
The replacement of $n$ in $\Delta^n_{\v k, \v q}$ with $n =1,
2, 3, 4$ and the proper choice of constant factors of $4$ or $8$ can
give us all kinds of physical spin susceptibility defined in Eq.~\eqref{eq:sus}.

\newpage
\noindent{\bf  RPA calculation with sublattice degree of freedom}
\\
In this Supplementary Information we present results for the magnetic nesting instabilities within the tight-binding random phase approximation(RPA) model including the sublattice and $J_{eff} =1/2$ degree of freedoms. We thus provide the sign and factor in front of $U$ for the RPA calculation of the spin susceptibility. With the final result in Eq.~\eqref{eqS:RPA} below, we can determine the critical value of $U$ when the denominator satisfies the divergence condition  for example, $1 = U \chi_0 ({\bf q})$ with the perfect nesting vector ${\bf q}$. The obtained critical values of $U$ as a function of the rotational angle $\theta$ enable us to determine the phase boundary of the magnetic phase diagram in the main text. Let us first write down spin susceptibility in AB-sublattice system.
\begin{align}
      \chi^{+-}_{AB} ({\bf q}, \tau) &= -\frac{1}{V} \sum_{{\bf p}, {\bf k}} \sum_{\sigma_1, \sigma_1'} \sum_{\sigma_2, \sigma_2'} \sum_{\alpha_1, \alpha_1'} \sum_{\beta_1, \beta_1'} \langle \mathbb{T}_{\tau} C^{\dag}_{{\bf k}, \alpha_1, \sigma_1} (\tau) \left( \sigma^+\right)_{\sigma_1, \sigma_1'} \left( s^z\right)_{\alpha_1, \alpha_1'} C_{{\bf k} + {\bf q}, \alpha_1', \sigma_1'} (\tau) \nonumber\\
&~~~~~~~~~~~~~~\times C^{\dag}_{{\bf p} + {\bf q}, \beta_1, \sigma_2} (0) \left( \sigma^-\right)_{\sigma_2, \sigma_2'}\left(s^z\right)_{\beta_1, \beta_1'} C_{{\bf p}, \beta_1', \sigma_2'} (0)\rangle \nonumber\\
&=\frac{1}{V} \sum_{{\bf p}, {\bf k}} \sum_{\sigma_1, \sigma_1'} \sum_{\sigma_2, \sigma_2'} \sum_{\alpha_1, \alpha_1'} \sum_{\beta_1, \beta_1'} \left( \sigma^+\right)_{\sigma_1, \sigma_1'} \left( s^z\right)_{\alpha_1, \alpha_1'} \left( \sigma^-\right)_{\sigma_2, \sigma_2'}\left(s^z\right)_{\beta_1, \beta_1'} \nonumber\\
&~~~~~~~~~~~~~~~\times \langle \mathbb{T}_{\tau} C_{{\bf p}, \beta_1', \sigma_2'} (0) C^{\dag}_{{\bf k}, \alpha_1, \sigma_1} (\tau) C_{{\bf k} + {\bf q}, \alpha_1', \sigma_1'} (\tau) C^{\dag}_{{\bf p} + {\bf q}, \beta_1, \sigma_2} (0) \rangle,
\end{align}
where $\alpha_1, \alpha_1', \beta_1, \beta_1',$ denote the sublattice indices and $\sigma_1, \sigma_1', \sigma_2, \sigma_2',$ the spin indices. Here $\tau$ serves as the imaginary time not sublattice index as in main text. We formulate the equation of motion for RPA susceptibility. The derivatives of $\Theta(\tau)$ function in the braket gives $\delta(\tau)$ function
\begin{equation}
  \delta(0) \langle C_{{\bf p}, \beta_1', \sigma_2'}(0) C^{\dag}_{{\bf k}, \alpha_1, \sigma_1} (0) C_{{\bf k}+{\bf q}, \alpha_1', \sigma_1'}(0) C^{\dag}_{{\bf p}+{\bf q}, \beta_1, \sigma_2}(0)\rangle = - \langle C^{\dag}_{{\bf k}, \alpha_1, \sigma_1} C_{{\bf k}+{\bf q}, \alpha_1', \sigma_1'} C^{\dag}_{{\bf p}+{\bf q}, \beta_1, \sigma_2} C_{{\bf p}, \beta_1', \sigma_2'}\rangle
  \label{eq2}
\end{equation}
Employing Wick's theorem, Eq.~\eqref{eq2} becomes
\begin{align}
  & - \langle C^{\dag}_{{\bf k}, \alpha_1, \sigma_1} C_{{\bf k}+{\bf q}, \alpha_1', \sigma_1'} \rangle \langle C^{\dag}_{{\bf p}+{\bf q}, \beta_1, \sigma_2} C_{{\bf p}, \beta_1', \sigma_2'} \rangle - \langle C^{\dag}_{{\bf k}, \alpha_1, \sigma_1} C_{{\bf p}, \beta_1', \sigma_2'} \rangle \langle C^{\dag}_{{\bf p}+{\bf q}, \beta_1, \sigma_2} C_{{\bf k}+{\bf q}, \alpha_1', \sigma_1'}\rangle \nonumber\\
  &~~~ + \langle C^{\dag}_{{\bf k}, \alpha_1, \sigma_1} C^{\dag}_{{\bf p}+{\bf q}, \beta_1, \sigma_2} \rangle \langle C_{{\bf k}+{\bf q}, \alpha_1', \sigma_1'} C_{{\bf p}, \beta_1', \sigma_2'} \rangle \nonumber\\
  & = - \langle C^{\dag}_{{\bf k}, \alpha_1, \sigma_1} C_{{\bf p}, \beta_1', \sigma_2'} \rangle \delta_{{\bf p}, {\bf k} } \delta_{\beta_1, \alpha_1'} \delta_{\sigma_2, \sigma_1'} + \langle C^{\dag}_{{\bf p}+{\bf q}, \beta_1, \sigma_2} C_{{\bf k}+{\bf q}, \alpha_1', \sigma_1'}\rangle \delta_{{\bf p}, {\bf k} } \delta_{\beta_1', \alpha_1} \delta_{\sigma_2', \sigma_1},
\label{eq3}
\end{align}
where only the second term in first line contributes to the last line accounting for matrix element of $\sigma^+$ and $\sigma^-$ and $\langle C^{\dag} C^{\dag}\rangle = \langle C C\rangle = 0$. Implementing $\sigma^+$, $\sigma^-$ and $s^z$ Eq.~\eqref{eq3} reads
\begin{align}
  & - \langle C^{\dag}_{{\bf k}, \alpha_1, \sigma_1} C_{{\bf p}, \beta_1', \sigma_2'} \rangle (\sigma^+)_{\sigma_1 \sigma_1'} (\sigma^-)_{\sigma_1' \sigma_2'}(s^z)_{\alpha_1 \alpha_1'} (s^z)_{\alpha_1' \beta_1'} \nonumber\\
  &  + \langle C^{\dag}_{{\bf p}+{\bf q}, \beta_1, \sigma_2} C_{{\bf k}+{\bf q}, \alpha_1', \sigma_1'}\rangle (\sigma^+)_{\sigma_1 \sigma_1'} (\sigma^-)_{\sigma_2 \sigma_1}(s^z)_{\alpha_1 \alpha_1'} (s^z)_{\beta_1 \alpha_1} \nonumber\\
  & = -\langle n_{{\bf k}, \alpha, \uparrow}\rangle + \langle n_{{\bf k} +{\bf q}, \alpha, \downarrow} \rangle,
\end{align}
where we reach the last line when the spin index summations are applied. Next, we should evaluate the $\frac{\partial}{\partial \tau} C^{\dag}_{{\bf k}, \alpha, \sigma} (\tau)$. The Hamiltonian is given by
\begin{align}
  H &= H_0 + H_{\textrm{int}} \nonumber\\
    &= \sum_{{\bf k'}, \alpha, \sigma} \varepsilon_{{\bf k'}} C^{\dag}_{{\bf k'}, \alpha, \sigma}C_{{\bf k'}, \alpha, \sigma} + \frac{U}{N}\sum_{{\bf p},{\bf l}, {\bf q_1}, \sigma, \sigma', b} C^{\dag}_{{\bf p}, b, \sigma} C^{\dag}_{{\bf l}, b, \sigma'} C_{{\bf l} +{\bf q_1}, b, \sigma'} C_{{\bf p}- {\bf q_1}, b, \sigma},
\end{align}
with $b$ the sublattice index. There is no inter-sublattice mixing term in Hubbard interaction due to the nature of on-site interaction.  We obtain
\begin{align}
&[H_0, C^{\dag}_{{\bf k}, \alpha_1, \sigma_1}] (\tau) = \varepsilon_{{\bf k}} C^{\dag}_{{\bf k}, \alpha_1, \sigma_1} (\tau),\nonumber\\
&[H_0, C_{{\bf k} + {\bf q}, \alpha_1', \sigma_1'}] (\tau) = - \varepsilon_{{\bf k}+{\bf q}} C_{{\bf k}+{\bf q}, \alpha_1', \sigma_1'} (\tau), \nonumber\\
&[H_{\textrm{int}}, C^{\dag}_{{\bf k}, \alpha_1, \sigma_1}] (\tau) = \frac{U}{N}\sum_{{\bf p},{\bf l}, {\bf q_1}, \sigma, \sigma', b}[C^{\dag}_{{\bf p}, b, \sigma} C^{\dag}_{{\bf l}, b, \sigma'} C_{{\bf l} +{\bf q_1}, b, \sigma'} C_{{\bf p}- {\bf q_1}, b, \sigma} , C^{\dag}_{{\bf k}, \alpha_1, \sigma_1} ](\tau),\nonumber\\
& = \frac{U}{N}\sum_{{\bf p},{\bf l}, {\bf q_1}, \sigma, \sigma', b}\left(C^{\dag}_{{\bf p}, b, \sigma} C^{\dag}_{{\bf l}, b, \sigma'} ( C_{{\bf l} +{\bf q_1}, b, \sigma'} C_{{\bf p}- {\bf q_1}, b, \sigma} C^{\dag}_{{\bf k}, \alpha_1, \sigma_1} - C^{\dag}_{{\bf k}, \alpha_1, \sigma_1} C_{{\bf l} +{\bf q_1}, b, \sigma'} C_{{\bf p}- {\bf q_1}, b, \sigma} ) \right) (\tau)\nonumber\\
& = \frac{U}{N}\sum_{{\bf p},{\bf l}, {\bf q_1}, \sigma, \sigma', b}\left(C^{\dag}_{{\bf p}, b, \sigma} C^{\dag}_{{\bf l}, b, \sigma'} ( C_{{\bf l} +{\bf q_1}, b, \sigma'} \delta_{{\bf p}- {\bf q_1}, {\bf k}} \delta_{b, \alpha_1} \delta_{\sigma, \sigma_1} - \delta_{{\bf l} + {\bf q_1}, {\bf k}} \delta_{b,\alpha_1} \delta_{\sigma_1, \sigma'} C_{{\bf p}- {\bf q_1}, b, \sigma} ) \right) (\tau)\nonumber \\
& = \frac{U}{N}\sum_{{\bf p},{\bf l}, {\bf q_1}, \sigma, \sigma', b}\left(C^{\dag}_{{\bf k} + {\bf q_1}, \alpha_1, \sigma_1} C^{\dag}_{{\bf l}, \alpha_1, \sigma'} C_{{\bf l} +{\bf q_1}, \alpha_1, \sigma'} -C^{\dag}_{{\bf p}, \alpha_1, \sigma} C^{\dag}_{{\bf k} - {\bf q_1}, \alpha_1, \sigma_1} C_{{\bf p}- {\bf q_1}, \alpha_1, \sigma} ) \right) (\tau)
\end{align}
\begin{align}
&[H_{\textrm{int}}, C_{{\bf k} + {\bf q}, \alpha_1', \sigma_1'}] (\tau) = \frac{U}{N}\sum_{{\bf p},{\bf l}, {\bf q_1}, \sigma, \sigma', b}[C^{\dag}_{{\bf p}, b, \sigma} C^{\dag}_{{\bf l}, b, \sigma'} C_{{\bf l} +{\bf q_1}, b, \sigma'} C_{{\bf p}- {\bf q_1}, b, \sigma} , C_{{\bf k} + {\bf q}, \alpha_1', \sigma_1'} ](\tau), \nonumber \\
&= \frac{U}{N}\sum_{{\bf p},{\bf l}, {\bf q_1}, \sigma, \sigma', b}\left( (C^{\dag}_{{\bf p}, b, \sigma} C^{\dag}_{{\bf l}, b, \sigma'} C_{{\bf k} + {\bf q}, \alpha_1', \sigma_1'} - C_{{\bf k} + {\bf q}, \alpha_1', \sigma_1'} C^{\dag}_{{\bf p}, b, \sigma} C^{\dag}_{{\bf l}, b, \sigma'} ) C_{{\bf l} +{\bf q_1}, b, \sigma'} C_{{\bf p}- {\bf q_1}, b, \sigma} \right) (\tau) \nonumber\\
&= \frac{U}{N}\sum_{{\bf p},{\bf l}, {\bf q_1}, \sigma, \sigma', b}\left( (C^{\dag}_{{\bf p}, b, \sigma} \delta_{{\bf l}, {\bf k} + {\bf q}} \delta_{b, \alpha_1'} \delta_{\sigma', \sigma_1'} - \delta_{{\bf k} + {\bf q}, {\bf p} } \delta_{b, \alpha_1'} \delta_{\sigma_1', \sigma} C^{\dag}_{{\bf l}, b, \sigma'} ) C_{{\bf l} +{\bf q_1}, b, \sigma'} C_{{\bf p}- {\bf q_1}, b, \sigma} \right) (\tau)\nonumber \\
&= \frac{U}{N}\sum\left(C^{\dag}_{{\bf p}, \alpha_1', \sigma} C_{{\bf k} + {\bf q} + {\bf q_1}, \alpha_1', \sigma_1'} C_{{\bf p} -{\bf q_1}, \alpha_1', \sigma} -C^{\dag}_{{\bf l}, \alpha_1', \sigma'} C_{{\bf l} + {\bf q_1}, \alpha_1', \sigma'} C_{{\bf k} + {\bf q}- {\bf q_1}, \alpha_1', \sigma_1'} ) \right) (\tau)
\end{align}
The third and fourth commutators gives $-\langle n_{{\bf k}, \alpha_1, \uparrow}\rangle $ and $\langle n_{{\bf k} + {\bf q}, \alpha_1', \downarrow}\rangle $ contributions. The resultant Dyson's equation is thus
\begin{equation} (i \omega_n - \varepsilon_{{\bf k}} + \varepsilon_{{\bf k}+ {\bf k}}) \chi_{AB}^{+-} ({\bf q}, i \omega_n) = (-\langle n_{{\bf k}, \alpha_1, \uparrow}\rangle + \langle n_{{\bf k} + {\bf q}, \alpha_1', \downarrow}\rangle ) \left(1 + U \chi_{AB}^{+-} ({\bf q}, i \omega_n) \right)
\end{equation}
So, we have
\begin{equation}
  \chi_{AB}^{+-} ({\bf q}, i \omega_n) = \frac{ (\chi_{AB}^{+-}({\bf q}, i \omega_n))_0 }{ 1- U (\chi_{AB}^{+-} ({\bf q}, i \omega_n))_0}.
  \label{eqS:RPA}
\end{equation}

\newpage
\noindent{\bf Tetrahedron methods in two-dimensional system}
\\
As noted in Eq.~\eqref{eqS:sus-final}, the spin susceptibility takes the following form
\begin{equation}
  \chi_{nn'}({\bf q}) = \frac{1}{V}\sum_{{\bf k}} \frac{1}{E_n({\bf k}) - E_{n'} ({\bf k}+{\bf q})}.
  \label{eq:example-sus}
\end{equation}
The susceptibility diverges when it satisfies the nesting condition $E_{n} ({\bf k}) = E_{n'} ({\bf k} + {\bf q})$. Here we provide the analytic expression for $\chi_{nn'}({\bf q})$ by using the tetrahedron methods. In the previous work by Rath and Freeman in 1975~\cite{tetrahedron}, the tetrahedron methods are subjected to the integral over the three-dimensional ${\bf k}$-space. A variant formula to two-dimensional ${\bf k}$-space seems obvious but it is worth clarifying explicit form. Let us choose the coordinates of the corners of triangle
\begin{equation}
  {\bf k}_1 = (0, 0),~{\bf k}_2 = (X_1, 0), ~{\bf k}_2 = (X_2, Y_2),
  \end{equation}
  and we define $V_i = E_{n'}({\bf k}_i+{\bf q})-E_n({\bf k}_i)$ where $i = 1,2,3$. We then expand the energy difference linearly
  \begin{equation}
    E_{n'}({\bf k} + {\bf q}) - E_{n} ({\bf k}) = A +  B x + Cy.
  \end{equation}
  Here the coefficients $A, B, C$ can be obtained from the energy difference at the corner of the triangle:
  \begin{equation}
A = V_1,~ A+B X_1 = V_2, A+BX_2+C Y_2 =V_3.
\end{equation}
The integral over the triangle can be written as
\begin{equation}
  \chi = \int^{Y_2}_{0}dy \left[ \int^{\frac{(X_2-X_1)}{Y_2}y+X_1}_{\frac{X_2}{Y_2}y} \frac{1}{A+Bx+Cy}\right].
\end{equation}
Basically, we assume $V_1 < V_2 < V_3$. 
For analytic expression, we have
\begin{equation}
  \chi  = \frac{V_1 \ln(|V_1|)}{(V_1 - V_2)(V_1 - V_3) } + \frac{V_2
  \ln(|V_2|)}{(V_3 - V_2)(V_1 - V_2) } + \frac{V_3 \ln(|V_3|)}{(V_3 -
  V_2)(V_3 - V_1) }.
\end{equation}
It holds true both for $V_1 , V_2, V_3 > 0$, $V_1 , V_2, V_3 < 0$.\\
We must carefully treat the above expression in the limit of several cases:
\begin{enumerate}
%%%%
\item[i)] $V_1 = V_2 = V_3, V_1 > 0$  or $V_1 < 0$\\
  \begin{equation}
     \chi = \frac{1}{2V_1}.
    \end{equation}
%%%%
  \item[ii)] $V_1 = V_2 = V_3 = 0$ \\
    \begin{equation}
      \chi =  0.
      \end{equation}
%%%%
    \item[iii)] $V_1 = V_2 \neq V_3 $, $V_1 \neq 0$, $V_3 \neq 0$, $V_1 > 0,
      V_3 > 0$ and $V_1 < 0, V_3 < 0$ \\
      \begin{equation}
        \chi =  \frac{V_1 - V_3 + V_3
      \ln(|\frac{V_3}{V_1}|)}{(V_1- V_3)^2}.
  \end{equation}
%%%%
   \item[iv-1)] $V_1 = V_2 \neq V_3 $, $V_1 \neq 0$, $V_3 = 0$, $V_1 > 0,
      V_1 < 0$ \\
      \begin{equation}
        \chi =  \frac{1}{V_1}.
  \end{equation}
  %%%%
  \item[iv-2)] $V_1 = V_2 \neq V_3 $, $V_1 = V_2 = 0$, $V_3 \neq 0$\\
      \begin{equation}
\chi =  0.
  \end{equation}
  %%%%
   \item[v)] $V_1 \neq V_2 = V_3 $, $V_1 \neq 0$, $V_3 \neq 0$, $V_1 > 0,
      V_3 > 0$ and $V_1 < 0, V_3 < 0$ \\
      \begin{equation}
        \chi =  \frac{-V_1 + V_3 + V_1
      \ln(|\frac{V_1}{V_3}|)}{(V_1- V_3)^2}.
  \end{equation}
  %%%%
   \item[vi-1)] $V_1 \neq V_2 = V_3 $, $V_1 = 0$, $V_3 \neq 0$, $V_3 > 0,
      V_3 < 0$ \\
      \begin{equation}
        \chi = \frac{1}{V_3}.
      \end{equation}
      %%%%
   \item[vi-2)] $V_1 \neq V_2 = V_3 $, $V_1 \neq 0$, $V_2 = V_3 =  0$ \\
      \begin{equation}
\chi =  0.
      \end{equation}
%%%%
   \item[vii)] $V_1 = 0,  V_2 \neq V_3 $, $V_2, V_3 > 0$ \\
      \begin{equation}
        \chi = \frac{\ln(|\frac{V_2}{V_3}|)}{V_2 - V_3}.
        \end{equation}
%%%%
   \item[viii)] $V_3 = 0,  V_1 \neq V_2 $, $V_1, V_2 < 0$ \\
      \begin{equation}
        \chi = \frac{\ln(|\frac{V_1}{V_2}|)}{V_1 - V_2}.
      \end{equation}
 %%%%     
  \item[ix)] $V_2 = 0,  V_1 \neq V_3 $, $V_1<0, V_3 > 0$ \\
      \begin{equation}
        \chi = \frac{\ln(|\frac{V_1}{V_3}|)}{V_1 - V_3}.
      \end{equation}
 %%%%
      
  \end{enumerate}
  With this exact form of susceptibility one can properly capture the logarithmically diverging feature as in Fig.~5{\bf d} in the maintext.

\newpage
\noindent{\bf AFM domain wall in-gap states}
\\
In this section, we analyze the in-gap states localized in antiferromagnetic domain wall of Sr$_2$IrO$_4$ system.

\section{The Full Hamiltonian}
The tight-binding (TB) Hamiltonian of a single-layer strontium iridate is given by
\begin{equation}
H(\textbf{k})=(\varepsilon_{2}(\textbf{k},\theta)+\varepsilon_{3}(\textbf{k},\theta))\sigma_{0}\tau_{0}+\varepsilon_{1}(\textbf{k},\theta)\sigma_{0}\tau_{x}+\varepsilon_{1d}(\textbf{k},\theta)\sigma_{z}\tau_{y},
\end{equation}
where
\begin{align*}
\varepsilon_{2}(\textbf{k},\theta)&=4t_{2}(\theta)\cos{k_{x}}\cos{k_{y}}\\
\varepsilon_{3}(\textbf{k},\theta)&=2t_{3}(\theta)(\cos{2k_{x}}+\cos{2k_{y}})\\
\varepsilon_{1}(\textbf{k},\theta)&=2t_{1}(\theta)(\cos{k_{x}}+\cos{k_{y}})\\
\varepsilon_{1d}(\textbf{k},\theta)&=2t_{1d}(\theta)(\cos{k_{x}}+\cos{k_{y}}).
\end{align*}

Or, rewriting the above equation in matrix form, we have
\begin{equation}
\begin{split}
&H(k_{x},k_{y})=(4t_{2}\cos{k_{x}}\cos{k_{y}}+2t_{3}(\cos{2k_{x}}+\cos{2k_{y}}))\sigma_{0}\tau_{0}+\\
&\begin{pmatrix}\!\!\!\!
0 \!\!\!\!\!\!\!\!&\!\!\!\!\!\!\!\! 2(t_{1}-i t_{1d})(\cos{k_{x}}+\cos{k_{y}}) \!\!\!\!\!\!\!\!&\!\!\!\!\!\!\!\! 0 \!\!\!\!\!\!\!\!&\!\!\!\!\!\!\!\! 0 \\
2(t_{1}+i t_{1d})(\cos{k_{x}}+\cos{k_{y}}) \!\!\!\!\!\!\!\!&\!\!\!\!\!\!\!\! 0 \!\!\!\!\!\!\!\!&\!\!\!\!\!\!\!\! 0 \!\!\!\!\!\!\!\!&\!\!\!\!\!\!\!\! 0 \\
0 \!\!\!\!\!\!\!\!&\!\!\!\!\!\!\!\! 0 \!\!\!\!\!\!\!\!&\!\!\!\!\!\!\!\! 0 \!\!\!\!\!\!\!\!&\!\!\!\!\!\!\!\! 2(t_{1}+i t_{1d})(\cos{k_{x}}+\cos{k_{y}}) \\
0 \!\!\!\!\!\!\!\!&\!\!\!\!\!\!\!\! 0 \!\!\!\!\!\!\!\!&\!\!\!\!\!\!\!\! 2(t_{1}-i t_{1d})(\cos{k_{x}}+\cos{k_{y}}) \!\!\!\!\!\!\!\!&\!\!\!\!\!\!\!\! 0
\!\!\!\!
\end{pmatrix}.
\end{split}
\end{equation}

When we introduce a magnetic ordering into the system, the Hamiltonian becomes
\begin{equation}
\begin{split}
&H(k_{x},k_{y})=(4t_{2}\cos{k_{x}}\cos{k_{y}}+2t_{3}(\cos{2k_{x}}+\cos{2k_{y}}))\sigma_{0}\tau_{0}+\\
&\begin{pmatrix}
0 \!\!\!\!\!\!\!\!&\!\!\!\!\!\!\!\! 2(t_{1}-i t_{1d})(\cos{k_{x}}+\cos{k_{y}}) \!\!\!\!\!\!\!\!&\!\!\!\!\!\!\!\! m_{x}^{A}-im_{y}^{A} \!\!\!\!\!\!\!\!&\!\!\!\!\!\!\!\! 0 \\
2(t_{1}+i t_{1d})(\cos{k_{x}}+\cos{k_{y}}) \!\!\!\!\!\!\!\!&\!\!\!\!\!\!\!\! 0 \!\!\!\!\!\!\!\!&\!\!\!\!\!\!\!\! 0 \!\!\!\!\!\!\!\!&\!\!\!\!\!\!\!\! m_{x}^{B}-im_{y}^{B} \\
m_{x}^{A}+im_{y}^{A} \!\!\!\!\!\!\!\!&\!\!\!\!\!\!\!\! 0 \!\!\!\!\!\!\!\!&\!\!\!\!\!\!\!\! 0 \!\!\!\!\!\!\!\!&\!\!\!\!\!\!\!\! 2(t_{1}+i t_{1d})(\cos{k_{x}}+\cos{k_{y}}) \\
0 \!\!\!\!\!\!\!\!&\!\!\!\!\!\!\!\! m_{x}^{B}+im_{y}^{B} \!\!\!\!\!\!\!\!&\!\!\!\!\!\!\!\! 2(t_{1}-i t_{1d})(\cos{k_{x}}+\cos{k_{y}}) \!\!\!\!\!\!\!\!&\!\!\!\!\!\!\!\! 0
\end{pmatrix}.
\end{split}
\end{equation}

For the sake of convenience in later calculations, we choose a new set of coordinates $(K_{X},K_{Y})=\frac{1}{\sqrt{2}}(k_{x}+k_{y},k_{y}-k_{x})$. Then the Hamiltonian reads
\begin{equation}
\begin{split}
&H(K_{X},K_{Y})=(2t_{2}(\cos{\sqrt{2}K_{X}}+\cos{\sqrt{2}K_{Y}})+4t_{3}\cos{\sqrt{2}K_{X}}\cos{\sqrt{2}K_{Y}})\sigma_{0}\tau_{0}+\\
&\begin{pmatrix}
0 & 4(t_{1}-i t_{1d})\frac{\cos{K_{X}}}{\sqrt{2}}\frac{\cos{K_{Y}}}{\sqrt{2}} & m_{x}^{A}-im_{y}^{A} & 0 \\
4(t_{1}+i t_{1d})\frac{\cos{K_{X}}}{\sqrt{2}}\frac{\cos{K_{Y}}}{\sqrt{2}} & 0 & 0 & m_{x}^{B}-im_{y}^{B} \\
m_{x}^{A}+im_{y}^{A} & 0 & 0 & 4(t_{1}+i t_{1d})\frac{\cos{K_{X}}}{\sqrt{2}}\frac{\cos{K_{Y}}}{\sqrt{2}} \\
0 & m_{x}^{B}+im_{y}^{B} & 4(t_{1}-i t_{1d})\frac{\cos{K_{X}}}{\sqrt{2}}\frac{\cos{K_{Y}}}{\sqrt{2}} & 0
\end{pmatrix}.
\end{split}
\end{equation}

\section{The Low Energy Effective Hamiltonian}
We already know that the band structure of the Hamiltonian (2) has a four-fold degenerate nodal line, close to the Fermi energy, along the Brillouin zone (BZ) boundary.
Since we are interested in the low energy physics near the Fermi level, we expand the Hamiltonian around a certain point on the BZ boundary $(K_{X0},K_{Y0})$.
As we set $(K_{X0},K_{Y0})=(0,\pi/\sqrt{2})$,
\begin{align}
%\begin{split}
&H(\delta K_{X},\delta K_{Y})+4t_{3}\sigma_{0}\tau_{0}=\nonumber\\
&\begin{pmatrix}
0 & -2\sqrt{2}(t_{1}-i t_{1d})\delta K_{Y} & m_{x}^{A}-im_{y}^{A} & 0 \\
-2\sqrt{2}(t_{1}+i t_{1d})\delta K_{Y} & 0 & 0 & m_{x}^{B}-im_{y}^{B} \\
m_{x}^{A}+im_{y}^{A} & 0 & 0 & -2\sqrt{2}(t_{1}+i t_{1d})\delta K_{Y} \\
0 & m_{x}^{B}+im_{y}^{B} & -2\sqrt{2}(t_{1}-i t_{1d})\delta K_{Y} & 0
\end{pmatrix}
+\mathcal{O}(\delta K^{2}).
%\end{split}
\end{align}
The $4t_{3}\sigma_{0}\tau_{0}$ term does nothing but just give a constant shift to the band structure, so we neglect it from now on.
Then, up to the first order of $\delta K_{i}$'s, the effective Hamiltonian is written as
\begin{equation}
H(\delta K_{Y})=
\begin{pmatrix}
0 & -2\sqrt{2}(t_{1}-i t_{1d})\delta K_{Y} & m_{x}^{A}-im_{y}^{A} & 0 \\
-2\sqrt{2}(t_{1}+i t_{1d})\delta K_{Y} & 0 & 0 & m_{x}^{B}-im_{y}^{B} \\
m_{x}^{A}+im_{y}^{A} & 0 & 0 & -2\sqrt{2}(t_{1}+i t_{1d})\delta K_{Y} \\
0 & m_{x}^{B}+im_{y}^{B} & -2\sqrt{2}(t_{1}-i t_{1d})\delta K_{Y} & 0
\end{pmatrix}.
\end{equation}

\section{A Single Domain}
To consider a single domain with the net ferromagnetic moment in $+Y$ direction, we put 
\begin{align}
m_{x}^{A}&=m\cos\alpha\nonumber\\
m_{y}^{A}&=m\sin\alpha\nonumber\\
m_{x}^{B}&=-m\sin\alpha\nonumber\\
m_{y}^{B}&=-m\cos\alpha,
\end{align}
where $m$ and $\alpha$ are positive real constants which denote the magnitude of the magnetic ordering and the angle between $\vec{m}^{A}$ and $x$-axis, respectively. The Hamiltonian with such configuration has several local symmetries: $G$, $C$, and $M$.
\begin{align}
G&=\frac{1}{\sqrt{2}}(\sigma_{x}-\sigma_{y})\tau_{x}\nonumber\\
C&=\frac{1}{\sqrt{2}}(\sigma_{x}+\sigma_{y})\tau_{y}\nonumber\\
M&=\sigma_{z}\tau_{z}
\end{align}
$H(\delta K_{Y})$ commutes with G and anticommutes with C and M. Using the following similarity transformiation $U$ that diagonalizes G ($UGU^{\dagger}=\textrm{diag}(-1,-1,1,1)$),
\begin{equation}
U=
\begin{pmatrix}
-\frac{1-i}{2} & 0 & 0 & \frac{1}{\sqrt{2}} \\
0 & -\frac{1-i}{2} &  \frac{1}{\sqrt{2}} & 0 \\
\frac{1-i}{2} & 0 & 0 &  \frac{1}{\sqrt{2}} \\
0 & \frac{1-i}{2} &  \frac{1}{\sqrt{2}} & 0
\end{pmatrix},
\end{equation}
we can block diagonalize $H(\delta K_{Y})$ into
\begin{equation}
\begin{split}
H'(\delta K_{Y})&=UH(\delta K_{Y})U^{\dagger}\\
&=
\begin{pmatrix}
H_{u}(\delta K_{Y}) & 0 \\
0 & H_{l}(\delta K_{Y}) \\
\end{pmatrix},
\end{split}
\end{equation}
where
\begin{equation}
H_{u}(\delta K_{Y})=
\begin{pmatrix}
0 & -\tilde{t}\delta K_{Y}-\tilde{m} \\
-\tilde{t}^{*}\delta K_{Y}-\tilde{m}^{*} & 0
\end{pmatrix}\>\>\textrm{and}\,\,
H_{l}(\delta K_{Y})=
\begin{pmatrix}
0 & -\tilde{t}\delta K_{Y}+\tilde{m} \\
-\tilde{t}^{*}\delta K_{Y}+\tilde{m}^{*} & 0
\end{pmatrix}.
\end{equation}
Here, $\tilde{t}$ and $\tilde{m}$ are defined as $\tilde{t}=2\sqrt{2}(t_{3}-i t_{4})$ and $\tilde{m}=me^{-i(\alpha+\frac{\pi}{4})}$.

Eigenvalues of each blocks are given by
\begin{align}
E_{u}(\delta K_{Y})&=\pm\left|\tilde{t}\delta K_{Y}+\tilde{m}\right|,\nonumber\\
E_{l}(\delta K_{Y})&=\pm\left|\tilde{t}\delta K_{Y}-\tilde{m}\right|.
\end{align}
When $\delta K_{Y}$ goes to zero, the eigenvalues become $\pm m$. Thus, we confirm that a gap with size $2m$ opens at the point $(K_{X0},K_{Y0})=(0,\pi/\sqrt{2})$ in the case of the single magnetic domain.

\section{A Domain Wall Along [110] Direction}
Now we think about a system with a domain wall that separates two domains with different net ferromagnetic moments: one in $+Y$ direction, and the other in $-Y$ direction.
We can get the Hamiltonian of such system by modifing the magnetic ordering used in the previous section. In this section, we investigate three different types of domain wall models: Smooth, N\'{e}el, and Bloch domain wall.

\subsection{Smooth Wall}
First, we consider the simplest model, in which the magnitude of the magnetic moments changes, but their directions stay still.
In such a \textit{smooth wall}, the magnitude of the magnetic moments is smoothly scaled down to zero in the transition region.
Multiplying $\tanh(\beta Y)$ to (S56), we have the smooth wall configuration with the domain wall liying in the $Y=0$ plane.
\begin{align}
m_{x}^{A}&=m\cos\alpha\tanh(\beta Y),\nonumber\\
m_{y}^{A}&=m\sin\alpha\tanh(\beta Y),\nonumber\\
m_{x}^{B}&=-m\sin\alpha\tanh(\beta Y),\nonumber\\
m_{y}^{B}&=-m\cos\alpha\tanh(\beta Y).
\end{align}
The role of $\tanh(\beta Y)$ here is to invert the magnetic moments as $Y$ changes its sign, where $\left|\beta\right|$ determines the stiffness of the domail wall profile.

Under the similarity transformation introduced in the section III, $M$ transforms into $M'=\sigma_{0}\tau_{z}$.
As $H$ anticommutes with $M$, $H'$ anticommutes with $M'$. Thus, both of the $2\times2$ block Hamiltonians anticommute with $\tau_{z}$.
Because $H_{u}$ and $H_{l}$ anticommute with $\tau_{z}$, if these block Hamiltonians have zero energy eigenstates, the zero-eigenstates should also be eigenstates of $\tau_{z}$, so they are of the form
$\begin{pmatrix}
f(Y) \\
0
\end{pmatrix}$, or
$\begin{pmatrix}
0 \\
g(Y)
\end{pmatrix}$.

In the presence of the domain wall, the periodicity of the system along the $Y$ direction is broken, and thus $K_{Y}$ is no more a good quantum number.
Therefore, we replace $\delta K_{Y}$ by $(-i\partial_{Y})$ to solve the Hamiltonian equation. Then we have
\begin{align}
H_{u,scale}(\partial_{Y})&=
\begin{pmatrix}
0 & i\tilde{t}\partial_{Y}-\tilde{m}\tanh(\beta Y) \\
i\tilde{t}^{*}\partial_{Y}-\tilde{m}^{*}\tanh(\beta Y) & 0
\end{pmatrix},\nonumber\\
H_{l,scale}(\partial_{Y})&=
\begin{pmatrix}
0 & i\tilde{t}\partial_{Y}+\tilde{m}\tanh(\beta Y) \\
i\tilde{t}^{*}\partial_{Y}+\tilde{m}^{*}\tanh(\beta Y) & 0
\end{pmatrix}.
\end{align}
For the upper block, $(i\tilde{t}^{*}\partial_{Y}-\tilde{m}^{*}\tanh(\beta Y))f_{u}(Y)=0$ gives a solution $f_{u}(Y)\sim\cosh(\beta Y)^{-i\tilde{m}^{*}/\beta\tilde{t}^{*}}$,
while $(i\tilde{t}\partial_{Y}-\tilde{m}\tanh(\beta Y))g_{u}(Y)=0$ gives a solution $g_{u}(Y)\sim\cosh(\beta Y)^{-i\tilde{m}/\beta\tilde{t}}$.
On the other hand, for the lower block, $(i\tilde{t}^{*}\partial_{Y}+\tilde{m}^{*}\tanh(\beta Y))f_{l}(Y)=0$ gives a solution $f_{l}(Y)\sim\cosh(\beta Y)^{i\tilde{m}^{*}/\beta\tilde{t}^{*}}$,
and $(i\tilde{t}\partial_{Y}+\tilde{m}\tanh(\beta Y))g_{l}(Y)=0$ gives a solution $g_{l}(Y)\sim\cosh(\beta  Y)^{i\tilde{m}/\beta\tilde{t}}$.
Substituting $\tilde{m}/\tilde{t}$ to $c$, we can simply write the solutions in the following form
\begin{align}
f_{u}(Y)&\sim\cosh(\beta Y)^{-ic^{*}/\beta},\nonumber\\
g_{u}(Y)&\sim\cosh(\beta Y)^{-ic/\beta},\nonumber\\
f_{l}(Y)&\sim\cosh(\beta Y)^{ic^{*}/\beta},\nonumber\\
g_{l}(Y)&\sim\cosh(\beta Y)^{ic/\beta}.
\end{align}
In general, only two of these solutions are physically allowed, since we must discard solutions whose norms diverge as $Y\rightarrow\pm\infty$.
Which solutions survive depends on the signs of $m$, $\alpha$, and $\beta$ (and on the magnitude of $\alpha$ as well).
\begin{align}
c=\frac{\tilde{m}}{\tilde{t}}&=\frac{me^{-i(\alpha+\frac{\pi}{4})}}{t_{3}-it_{4}}\nonumber\\
&=\frac{me^{-i(\alpha+\frac{\pi}{4})}}{\left|\tilde{t}\right|e^{-i\phi}}\nonumber\\
&=\frac{m}{\left|\tilde{t}\right|}e^{-i(\alpha+\frac{\pi}{4}-\phi)}
\end{align}
where $\phi=\tan^{-1}(t_{4}/t_{3})$, $\left|f_{u}(Y)\right|^{2}=f_{u}^{*}(Y)f_{u}(Y)$ becomes
\begin{align}
\left|f_{u}(Y)\right|^{2}&\sim\cosh(\beta Y)^{-i\frac{m}{\beta\left|\tilde{t}\right|}(e^{i(\alpha+\frac{\pi}{4}-\phi)}-e^{-i(\alpha+\frac{\pi}{4}-\phi)})}\nonumber\\
&\sim\cosh(\beta Y)^{\frac{2m}{\beta\left|\tilde{t}\right|}\sin(\alpha+\frac{\pi}{4}-\phi)}.
\end{align}
Repeating the same calculation for the other solutions, we are left with
\begin{align}
\left|f_{u}(Y)\right|^{2}&\sim\cosh(\beta Y)^{\frac{2m}{\beta\left|\tilde{t}\right|}\sin(\alpha+\frac{\pi}{4}-\phi)},\nonumber\\
\left|g_{u}(Y)\right|^{2}&\sim\cosh(\beta Y)^{-\frac{2m}{\beta\left|\tilde{t}\right|}\sin(\alpha+\frac{\pi}{4}-\phi)},\nonumber\\
\left|f_{l}(Y)\right|^{2}&\sim\cosh(\beta Y)^{-\frac{2m}{\beta\left|\tilde{t}\right|}\sin(\alpha+\frac{\pi}{4}-\phi)},\nonumber\\
\left|g_{l}(Y)\right|^{2}&\sim\cosh(\beta Y)^{\frac{2m}{\beta\left|\tilde{t}\right|}\sin(\alpha+\frac{\pi}{4}-\phi)}.
\end{align}
Eq. (S67) states that if $\frac{2m}{\beta\left|\tilde{t}\right|}<0$, the two valid solutions would be $f_{u}$ and $g_{l}$, while if $\frac{2m}{\beta\left|\tilde{t}\right|}<0$ the valid solutions would be $g_{u}$ and $f_{l}$.
For example, when $m$ and $\beta$ are given to be positive, and $(\phi-\alpha)<\pi/4$, $\left|f_{u}(Y)\right|^{2}$ and $\left|g_{l}(Y)\right|^{2}$ have positive exponents, so they are unphysical.
Meanwhile, $\left|g_{u}(Y)\right|^{2}$ and $\left|f_{l}(Y)\right|^{2}$ have negative exponents, therefore $g_{u}$ and $f_{l}$ are the final solutions that we have been seeking for.
\begin{figure}[htbp]
\begin{center}
    \includegraphics[scale=0.7]{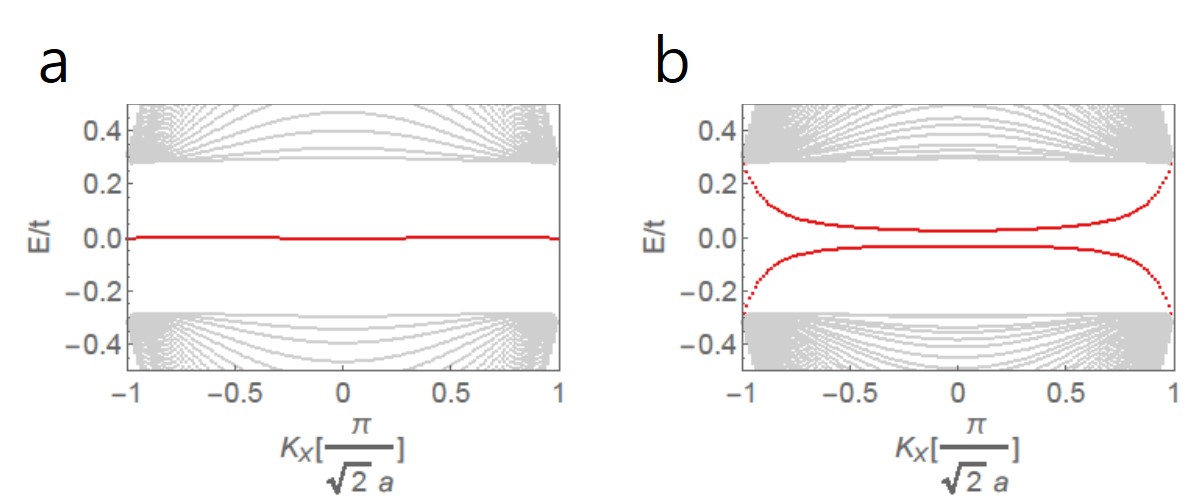}
    \caption{Dispersion of domain wall states of the smooth wall (a) with and (b) without zero magnetic moment atoms at the domain wall position.}\label{fig:label}
\end{center}
\end{figure}

However, numerical calculation for a finite size system does not always give the zero energy eigenstates. Only when the system has atoms located exactly on the domain wall thus there is a line of atoms with zero magentic moment on the $Y=0$ plane, \textit{i.e.} $n_{DW}=0$, the doubly degenerate zero mode appears, and otherwise, the domain wall states are gapped. It is similar to a situation that happens in the case of Su-Schrieffer–Heeger (SSH) model; a domain wall in the SSH model exhibits zero modes if there is an atomic site right on the domain wall, but does not if a bond is located on the domain wall instead of an atomic site. 

\subsection{N\'{e}el Wall}
When magnetic moments rotate around an axis parallel to a domain wall plane in the transition region, it is called a N\'{e}el wall.
The domain wall configuration for the N\'{e}el wall is defined by
\begin{align}
\vec{m}^{A}&=m(\cos\alpha\tanh(\beta Y)+\sin\alpha\,\textrm{sech}(\beta Y),-\cos\alpha\,\textrm{sech}(\beta Y)+\sin\alpha\tanh(\beta Y),0),\nonumber\\
\vec{m}^{B}&=-m(\sin\alpha\tanh(\beta Y)+\cos\alpha\,\textrm{sech}(\beta Y),-\sin\alpha\,\textrm{sech}(\beta Y)+\cos\alpha\tanh(\beta Y),0).
\end{align}
Then the transformed effective Hamiltonian becomes
\begin{align}
&H'_{Neel}(\delta K_{Y})=\nonumber\\
&\begin{pmatrix}
0 & -\tilde{t}\delta K_{Y}-\tilde{m}\tanh(\beta Y) & 0 & -i\tilde{m}\,\textrm{sech}(\beta Y) \\
-\tilde{t}^{*}\delta K_{Y}-\tilde{m}^{*}\tanh(\beta Y) & 0 & -i\tilde{m}^{*}\,\textrm{sech}(\beta Y) & 0 \\
0 & i\tilde{m}\,\textrm{sech}(\beta Y) & 0 & -\tilde{t}\delta K_{Y}+\tilde{m}\tanh(\beta Y) \\
i\tilde{m}^{*}\,\textrm{sech}(\beta Y) & 0 & -\tilde{t}^{*}\delta K_{Y}+\tilde{m}^{*}\tanh(\beta Y) & 0
\end{pmatrix}.
\end{align}
Unlike $H_{u,scale}$ and $H_{l,scale}$ which were totally decoupled, it is obvious that $H_{u,Neel}$ and $H_{l,Neel}$ are coupled to each other. It means that the two domain wall state are mixed and a gap opens.
\begin{figure}[htbp]
\begin{center}
    \includegraphics[scale=0.7]{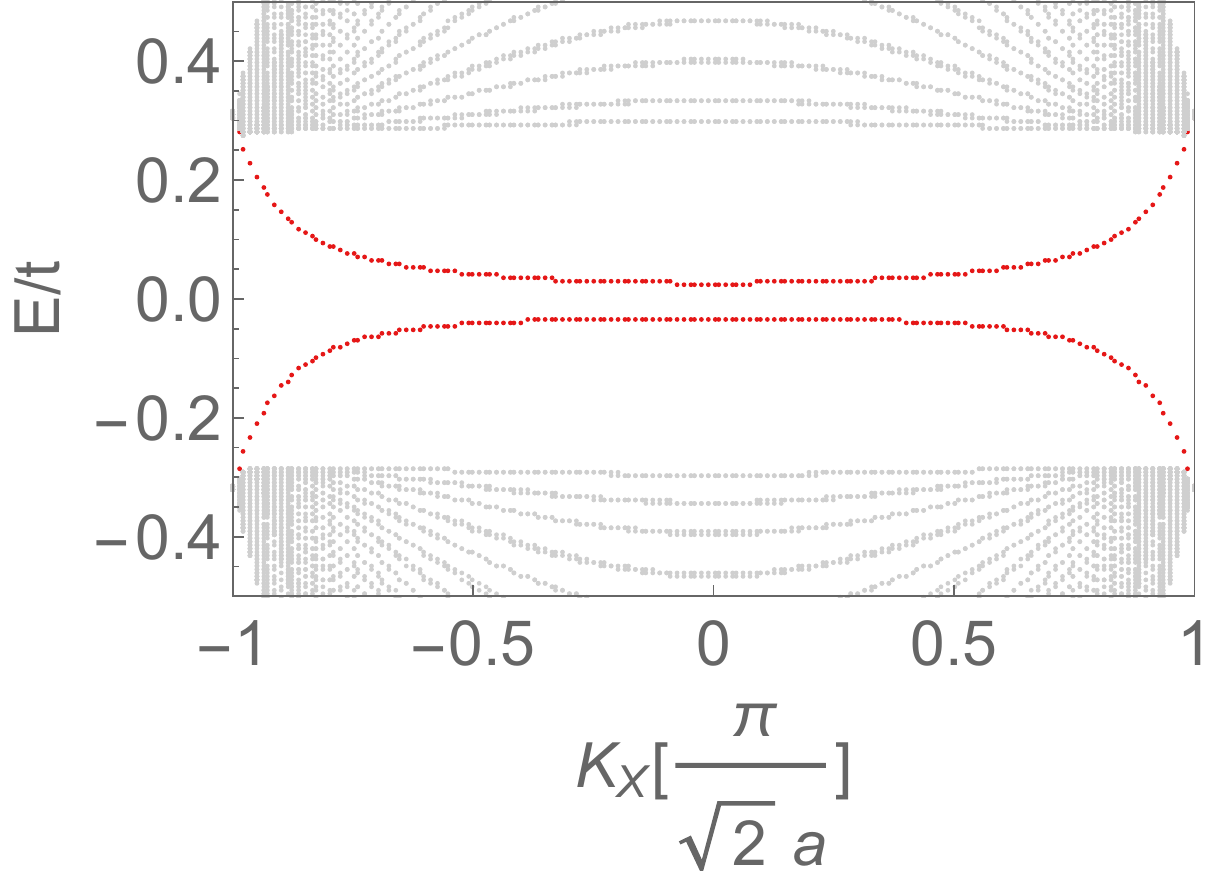}
    \caption{Dispersion of domain wall states of the N\'{e}el wall.}\label{fig:label}
\end{center}
\end{figure}

\subsection{Bloch Wall}
When magnetic moments rotate around an axis perpendicular to the domain wall plane in the transition region, it is called a Bloch wall.
The domain wall configuration for the Bloch wall is defined by
\begin{align}
\vec{m}^{A}&=m(\cos\alpha\tanh(\beta Y),\sin\alpha\tanh(\beta Y),\textrm{sech}(\beta Y)),\nonumber\\
\vec{m}^{B}&=-m(\sin\alpha\tanh(\beta Y),\cos\alpha\tanh(\beta Y),\textrm{sech}(\beta Y)).
\end{align}
Then the transformed effective Hamiltonian becomes
\begin{align}
&H'_{Bloch}(\delta K_{Y})=\nonumber\\
&\begin{pmatrix}
m\,\textrm{sech}(\beta Y) & -\tilde{t}\delta K_{Y}-\tilde{m}\tanh(\beta Y) & 0 & 0 \\
-\tilde{t}^{*}\delta K_{Y}-\tilde{m}^{*}\tanh(\beta Y) & -m\,\textrm{sech}(\beta Y) & 0 & 0 \\
0 & 0 & m\,\textrm{sech}(\beta Y) & -\tilde{t}\delta K_{Y}+\tilde{m}\tanh(\beta Y) \\
0 & 0 & -\tilde{t}^{*}\delta K_{Y}+\tilde{m}^{*}\tanh(\beta Y) & -m\,\textrm{sech}(\beta Y)
\end{pmatrix}.
\end{align}
This time, $H'_{u/l,Bloch}$ no more anti-commutes with $\tau_{z}$, so the zero modes do not exsist.
\begin{figure}[htbp]
\begin{center}
    \includegraphics[scale=0.7]{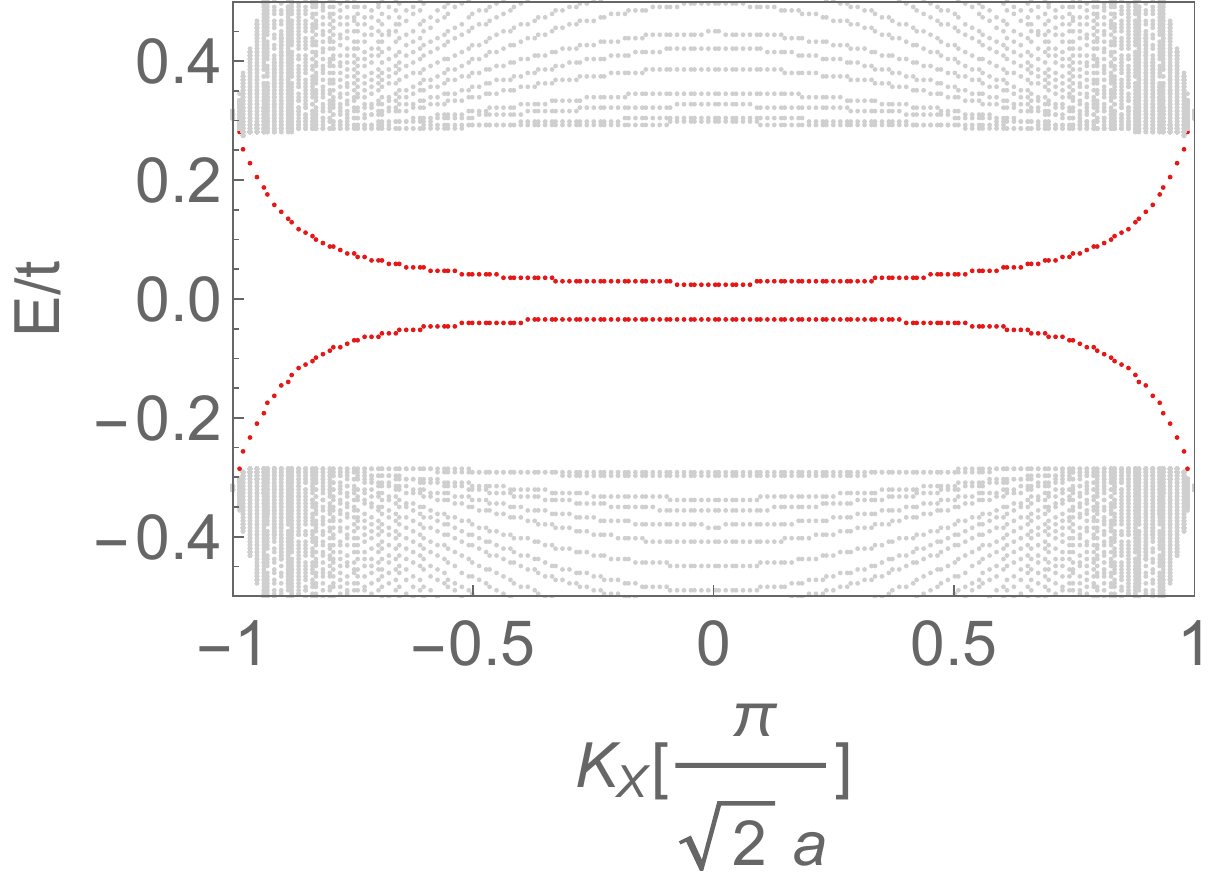}
    \caption{Dispersion of domain wall states of the Bloch wall.}\label{fig:label}
\end{center}
\end{figure}

\end{document}